\documentclass[%
 reprint,
superscriptaddress,
 amsmath,amssymb,
 aps,
]{revtex4-2}

\usepackage{graphicx}
\usepackage{float}
\usepackage{multirow}
\usepackage{dcolumn}
\usepackage{bm}
\usepackage[colorlinks,linkcolor=black,anchorcolor=black,citecolor=black]{hyperref}

\begin{document}

\preprint{APS/123-QED}

\title{Tunable Coupling Architectures with Capacitively Connecting Pads for Large-Scale Superconducting Multi-Qubit Processors}

\author{Gui-Han Liang}
\thanks{These authors contributed equally to this work.}
\affiliation{Institute of Physics, Chinese Academy of Sciences, Beijing 100190, China}
\affiliation{School of Physical Sciences, University of Chinese Academy of Sciences, Beijing 100049, China}

\author{Xiao-Hui Song}
\thanks{These authors contributed equally to this work.}
\affiliation{Institute of Physics, Chinese Academy of Sciences, Beijing 100190, China}
\affiliation{School of Physical Sciences, University of Chinese Academy of Sciences, Beijing 100049, China}
\affiliation{Beijing Academy of Quantum Information Sciences, Beijing 100193, China}
\affiliation{Hefei National Laboratory, Hefei 230088, China.}
\affiliation{CAS Center for Excellence in Topological Quantum Computation, University of Chinese Academy of Sciences, Beijing 100190, China}

\author{Cheng-Lin Deng}
\thanks{These authors contributed equally to this work.}
\affiliation{Institute of Physics, Chinese Academy of Sciences, Beijing 100190, China}
\affiliation{School of Physical Sciences, University of Chinese Academy of Sciences, Beijing 100049, China}

\author{Xu-Yang Gu}
\affiliation{Institute of Physics, Chinese Academy of Sciences, Beijing 100190, China}
\affiliation{School of Physical Sciences, University of Chinese Academy of Sciences, Beijing 100049, China}

\author{Yu Yan}
\affiliation{Institute of Physics, Chinese Academy of Sciences, Beijing 100190, China}
\affiliation{School of Physical Sciences, University of Chinese Academy of Sciences, Beijing 100049, China}

\author{Zheng-Yang Mei}
\affiliation{Institute of Physics, Chinese Academy of Sciences, Beijing 100190, China}
\affiliation{School of Physical Sciences, University of Chinese Academy of Sciences, Beijing 100049, China}

\author{Si-Lu Zhao}
\affiliation{Institute of Physics, Chinese Academy of Sciences, Beijing 100190, China}
\affiliation{School of Physical Sciences, University of Chinese Academy of Sciences, Beijing 100049, China}

\author{Yi-Zhou Bu}
\affiliation{Institute of Physics, Chinese Academy of Sciences, Beijing 100190, China}
\affiliation{School of Physical Sciences, University of Chinese Academy of Sciences, Beijing 100049, China}

\author{Yong-Xi Xiao}
\affiliation{Institute of Physics, Chinese Academy of Sciences, Beijing 100190, China}
\affiliation{School of Physical Sciences, University of Chinese Academy of Sciences, Beijing 100049, China}

\author{Yi-Han Yu}
\affiliation{Institute of Physics, Chinese Academy of Sciences, Beijing 100190, China}
\affiliation{School of Physical Sciences, University of Chinese Academy of Sciences, Beijing 100049, China}

\author{Ming-Chuan Wang}
\affiliation{Institute of Physics, Chinese Academy of Sciences, Beijing 100190, China}
\affiliation{School of Physical Sciences, University of Chinese Academy of Sciences, Beijing 100049, China}

\author{Tong Liu}
\affiliation{Institute of Physics, Chinese Academy of Sciences, Beijing 100190, China}
\affiliation{School of Physical Sciences, University of Chinese Academy of Sciences, Beijing 100049, China}

\author{Yun-Hao Shi}
\affiliation{Institute of Physics, Chinese Academy of Sciences, Beijing 100190, China}
\affiliation{School of Physical Sciences, University of Chinese Academy of Sciences, Beijing 100049, China}

\author{He Zhang}
\affiliation{Institute of Physics, Chinese Academy of Sciences, Beijing 100190, China}
\affiliation{School of Physical Sciences, University of Chinese Academy of Sciences, Beijing 100049, China}

\author{Xiang Li}
\affiliation{Institute of Physics, Chinese Academy of Sciences, Beijing 100190, China}
\affiliation{School of Physical Sciences, University of Chinese Academy of Sciences, Beijing 100049, China}

\author{Li Li}
\affiliation{Institute of Physics, Chinese Academy of Sciences, Beijing 100190, China}
\affiliation{School of Physical Sciences, University of Chinese Academy of Sciences, Beijing 100049, China}

\author{Jing-Zhe Wang}
\affiliation{Institute of Physics, Chinese Academy of Sciences, Beijing 100190, China}
\affiliation{School of Physical Sciences, University of Chinese Academy of Sciences, Beijing 100049, China}

\author{Ye Tian}
\affiliation{Institute of Physics, Chinese Academy of Sciences, Beijing 100190, China}
\affiliation{School of Physical Sciences, University of Chinese Academy of Sciences, Beijing 100049, China}

\author{Shi-Ping Zhao}
\affiliation{Institute of Physics, Chinese Academy of Sciences, Beijing 100190, China}
\affiliation{School of Physical Sciences, University of Chinese Academy of Sciences, Beijing 100049, China}

\author{Kai Xu}
\email{kaixu@iphy.ac.cn}
\affiliation{Institute of Physics, Chinese Academy of Sciences, Beijing 100190, China}
\affiliation{School of Physical Sciences, University of Chinese Academy of Sciences, Beijing 100049, China}
\affiliation{Beijing Academy of Quantum Information Sciences, Beijing 100193, China}
\affiliation{Hefei National Laboratory, Hefei 230088, China.}
\affiliation{CAS Center for Excellence in Topological Quantum Computation, University of Chinese Academy of Sciences, Beijing 100190, China}

\author{Heng Fan}
\email{hfan@iphy.ac.cn}
\affiliation{Institute of Physics, Chinese Academy of Sciences, Beijing 100190, China}
\affiliation{School of Physical Sciences, University of Chinese Academy of Sciences, Beijing 100049, China}
\affiliation{Beijing Academy of Quantum Information Sciences, Beijing 100193, China}
\affiliation{Hefei National Laboratory, Hefei 230088, China.}
\affiliation{CAS Center for Excellence in Topological Quantum Computation, University of Chinese Academy of Sciences, Beijing 100190, China}

\author{Zhong-Cheng Xiang}
\email{zcxiang@iphy.ac.cn}
\affiliation{Institute of Physics, Chinese Academy of Sciences, Beijing 100190, China}
\affiliation{School of Physical Sciences, University of Chinese Academy of Sciences, Beijing 100049, China}
\affiliation{Beijing Academy of Quantum Information Sciences, Beijing 100193, China}
\affiliation{Hefei National Laboratory, Hefei 230088, China.}
\affiliation{CAS Center for Excellence in Topological Quantum Computation, University of Chinese Academy of Sciences, Beijing 100190, China}

\author{Dong-Ning Zheng}
\email{dzheng@iphy.ac.cn}
\affiliation{Institute of Physics, Chinese Academy of Sciences, Beijing 100190, China}
\affiliation{School of Physical Sciences, University of Chinese Academy of Sciences, Beijing 100049, China}
\affiliation{Beijing Academy of Quantum Information Sciences, Beijing 100193, China}
\affiliation{Hefei National Laboratory, Hefei 230088, China.}
\affiliation{CAS Center for Excellence in Topological Quantum Computation, University of Chinese Academy of Sciences, Beijing 100190, China}

\date{\today}

\begin{abstract}
  We have proposed and experimentally verified a tunable inter-qubit coupling scheme for large-scale integration of superconducting qubits. The key feature of the scheme is the insertion of connecting pads between qubit and tunable coupling element. In such a way, the distance between two qubits can be increased considerably to a few millimeters, leaving enough space for arranging control lines, readout resonators and other necessary structures. The increased inter-qubit distance provides more wiring space for flip-chip process and reduces crosstalk between qubits and from control lines to qubits. We use the term Tunable Coupler with Capacitively Connecting Pad (TCCP) to name the tunable coupling part that consists of a transmon coupler and capacitively connecting pads. With the different placement of connecting pads, different TCCP architectures can be realized. We have designed and fabricated a few multi-qubit devices in which TCCP is used for coupling. The measured results show that the performance of the qubits coupled by the TCCP, such as $T_1$ and $T_2$, was similar to that of the traditional transmon qubits without TCCP. Meanwhile, our TCCP also exhibited a wide tunable range of the effective coupling strength and a low residual ZZ interaction between the qubits by properly tuning the parameters on the design. Finally, we successfully implemented an adiabatic CZ gate with TCCP. Furthermore, by introducing TCCP, we also discuss the realization of the flip-chip process and tunable coupling qubits between different chips.  
\end{abstract}

\maketitle


\section{Introduction \label{sec:1}}

The superconducting quantum processors have made significant progress over the past two decades through successive improvements in materials, device design and architecture, and fabrication process \cite{OliverReview,WallraffReview,SchoelkopfReview,IBMReview}. To perform various and complex quantum tasks and, particularly, fault-tolerant quantum computation, it is necessary to develop superconducting quantum processors with large number of qubits and with outstanding performance \cite{30bitTunableCavity,Sycamore,Zheda36bit,Zheda68bit,Zuchongzhi,Zuchongzhi2-2,20QubitFull,30QubitTizi,43Qubit}. A key element in multi-qubit devices is tunable coupler that is able to turn on/off the coupling between two adjacent qubits \cite{YanFeiTunable,Mediated,Parametric-Resonance,AntiFloating,FloatingCoupler,DoubleTransmonCoupler,LikeBCC}.

Transmon qubits are widely used in the superconducting quantum processors because of its long decoherence time. A transmon qubit can be regarded as a nonlinear LC oscillator, making it natural to couple two qubits capacitively or inductively. Tunable coupling schemes based on capacitive \cite{YanFeiTunable} and inductive \cite{Gmon} coupling have been demonstrated and implemented in large-scale multi-qubit processors. In these schemes, a coupling-off point can be realized and high qubit decoherence time can be maintained. A widely used scheme based on capacitive coupling with a high on/off ratio has been proposed \cite{YanFeiTunable}. However, in this design, the tunable coupler is a grounded transmon, and a large enough direct capacitive coupling between the qubits is required in the implementation of the tunable coupler. This could limit the circuit design because the qubits cannot be placed too far apart, making it difficult to provide adequate space for arranging readout resonators, control lines, airbridges, Purcell filters, and other necessary structures.

In another work \cite{FloatingCoupler}, the authors propose a floating tunable coupler that can turn off the effective qubit-qubit coupling without direct capacitive coupling between the qubits. However, stretching the architecture of the floating tunable coupler or qubits to obtain adequate wiring space may not only degrade qubit performance, resulting in reduced coherence time $T_1$ \cite{T1SurfaceLoss,Kelly,SurfaceLoss}, but also increase the design complexity of the qubit charge energy $E_C$ in the flip-chip process.

To address these issues, we propose a tunable coupling scheme that can increase the distance between qubits. We call it Tunable Coupler with Capacitively Connecting Pad (TCCP). The TCCP architectures consist of a grounded coupler and the capacitively connecting pads. These architectures can achieve a high on/off ratio and do not require direct capacitive coupling between qubits. With TCCP, qubits and tunable coupler can be designed with relatively small size, which can more effectively reduce parasitic capacitance caused by other layer along the stacking direction in the flip-chip process. In addition, the TCCP architectures can provide a connection form with tunable coupling between chips. Moreover, we experimentally fabricated the superconducting quantum processors with the TCCP architectures, demonstrating fine coherence time $T_1$ (average 24$\mu$s) of qubits and fidelity of controlled-Z (CZ) gate as high as 96.2$\%$. While writing up this manuscript, we found a reported work with a similar design \cite{LikeBCC,Patent}, but it also used floating tunable coupler. 

\section{Principle of TCCP Architectures \label{sec:2}}

\begin{figure*}[htb]
  \center{\includegraphics[width=17cm]  {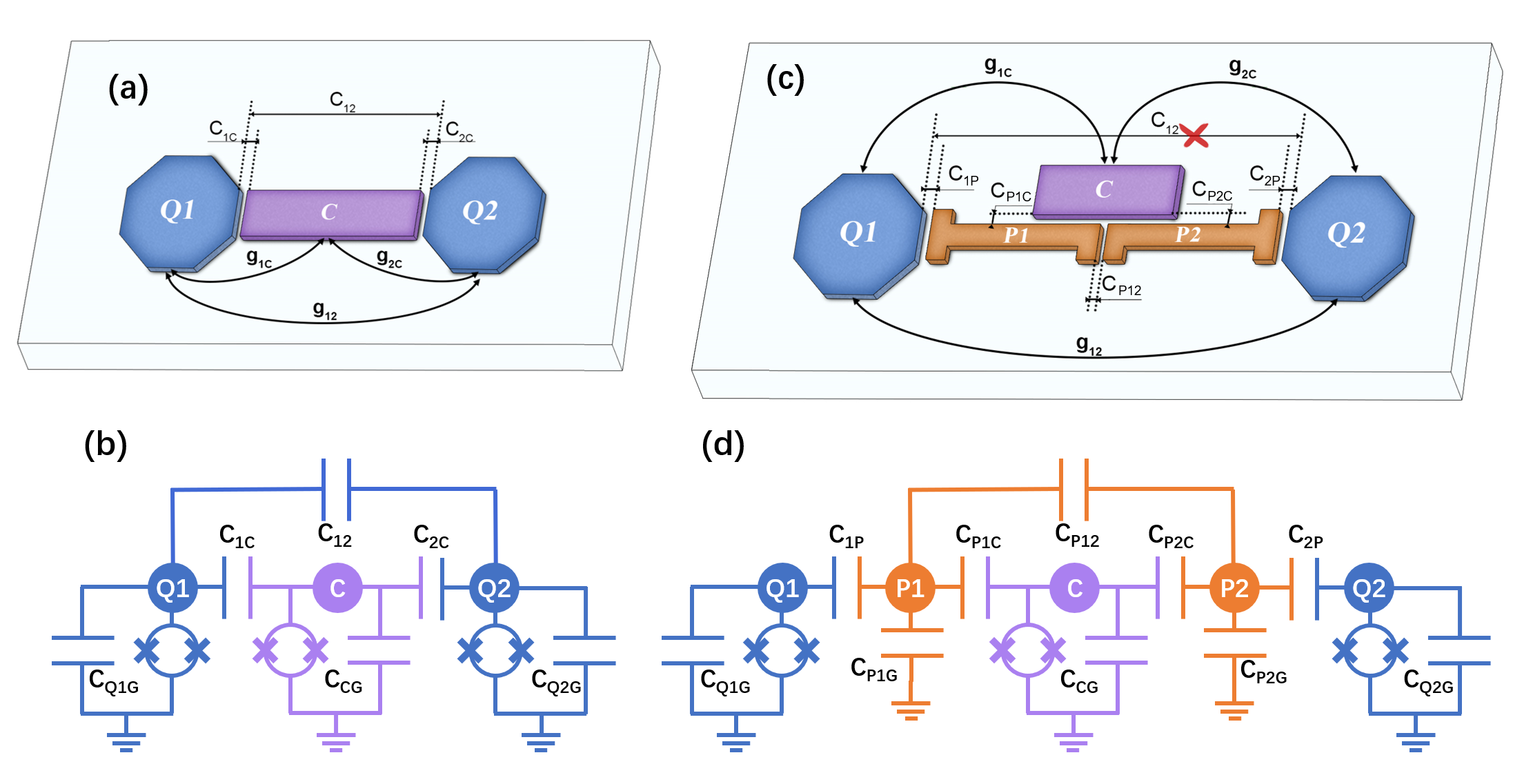}}
  \caption{(a) The schematic diagram of a typical tunable coupling scheme, which features a tunable coupler sandwiched between two qubits. It depicts the nearest-neighbor capacitances and couplings between the qubits and the tunable coupler. (b) The lumped-element circuit model of (a). (c) The schematic diagram of the TCCP architecture with two connecting pads. It displays the capacitances and couplings between the nearest-neighbor pads, while the capacitance between two qubits is negligible. (d) The lumped-element circuit model of (c).}
  \label{fig:1}
\end{figure*}

Before describing the principle of the TCCP architectures, we recall how a widely used design works. FIG.\ref{fig:1}(a) shows a typical tunable coupling scheme proposed by Fei Yan \cite{YanFeiTunable} where two qubits are coupled via a tunable coupler in form of capacitance. By analyzing and calculating its lumped circuit model shown in FIG.\ref{fig:1}(b), the effective coupling strength between the qubits is obtained:
\begin{equation}
    g_{\text{eff}}=g_{12}-\frac{g_{1\text{C}}g_{2\text{C}}}{2}\sum_{k=1}^2\left(\frac{1}{\Delta_k}+\frac{1}{\Sigma_k}\right) \label{equ:1}
\end{equation}
where $g_{\text{eff}}$ is the indirect virtual exchange coupling strength between the qubits via the tunable coupler. $g_{12}$ is the direct coupling strength between the qubits. $g_{k\text{C}}\ (k=1,2)$ are the direct coupling strengths between the qubits and the tunable coupler. $\omega_{\text{C}}$ and $\omega_k$ are the frequencies of the tunable coupler and the qubits, respectively, and $\Delta_k=\omega_{\text{C}}-\omega_k$, $\Sigma_k=\omega_{\text{C}}+\omega_k$. The effective coupling strength $g_{\text{eff}}$ can be tuned from negative to positive by changing the frequency of the tunable coupler. In the case of the tunable coupler realized by a grounded transmon, the direct capacitance $C_{12}\propto g_{12}$ plays a crucial role in canceling the effective coupling between the two qubits at a particular frequency of the tunable coupler when one qubit is brought closer to the other. However, the large wiring space required will be difficult to achieve, because the qubits cannot be placed far apart.

To solve this problem, we introduce TCCP architectures by inserting connecting pads between the qubits and the tunable coupler. FIG.\ref{fig:1}(c) shows one of such TCCP architectures with the arrangement of grounded qubit - connecting pad - grounded coupler - connecting pad - grounded qubit. FIG.\ref{fig:1}(d) gives its lumped-element circuit model. The Hamiltonian describing the system is derived in Appendix \ref{sec:A}. Notably, the flux coordinates $\Phi_{\text{P}1}$ and $\Phi_{\text{P}2}$ related to the connecting pads do not create any phase difference across the Josephson junction, and the modes represented by $\Phi_{\text{P}1}$ and $\Phi_{\text{P}2}$ are “free particles” rather than harmonic oscillators due to the vanishing of their spring constants and the absence of inductances associated with these modes. Therefore, they are not the qubits modes of interest and do not need to be included in the calculations \cite{Parametric-Resonance}. We can model this TCCP architecture readily using the Hamiltonian given by:
\begin{widetext}
\begin{equation}
  \hat{H}=\sum_{j=1,\text{C},2}\left[\omega_j+\frac{E_{Cj}}{2}\left(1-\frac{5\xi_j}{18}\right)-\frac{E_{Cj}}{2}\left(1-\frac{\xi_j}{6}\right)\hat{a}_j^\dagger\hat{a}_j\right]\hat{a}_j^\dagger\hat{a}_j+\sum_{k=1,2}g_{j\text{C}}(\hat{a}_k^\dagger-\hat{a}_k)(\hat{a}_\text{C}^\dagger-\hat{a}_\text{C})+g_{12}(\hat{a}_1^\dagger-\hat{a}_1)(\hat{a}_2^\dagger-\hat{a}_2)
  \label{equ:2}
\end{equation}
\end{widetext}
where $\omega_j/2\pi$, $E_{Cj}$, and $E_{Jj}$ are the frequencies, charging energies, and Josephson energies of qubits $(j=1,2)$ and tunable coupler $(j=\text{C})$, respectively. $\xi_j=\sqrt{2E_{Cj}/E_{Jj}}$ are the sixth-order correction, and $\hat{a}_j\ (\hat{a}_j^\dagger)$ are the annihilation (creation) operators. The direct qubit-coupler and qubit-qubit coupling strengths are:
\begin{subequations}
    \begin{align}
        g_{k\text{C}}&=\frac{E_{k\text{C}}}{\sqrt{2}}\left(\frac{E_{Jk}E_{J\text{C}}}{E_{Ck}E_{C\text{C}}}\right)^{\frac14}\left[1-\frac18(\xi_k+\xi_\text{C})\right],\ k\in\{1,2\}\\
        g_{12}&=\frac{E_{12}}{\sqrt{2}}\left(\frac{E_{J1}E_{J2}}{E_{C1}E_{C2}}\right)^{\frac14}\left[1-\frac18(\xi_1+\xi_2)\right]
    \end{align}\label{equ:3}
\end{subequations}
where $E_{k\text{C}}$ and $E_{12}$ are the direct qubit-coupler and qubit-qubit coupling charging energies. The effective qubit-qubit Hamiltonian can be obtained by approximating the qubits and the tunable coupler by their lowest two energy levels and applying a second-order Schrieffer-Wolff transformation \cite{SWT}. The resulting effective Hamiltonian can be expressed as:
\begin{subequations}
    \begin{align}
        \hat{H}_{\text{eff}}&=\sum_{k=1}^2\left(-\frac12\omega_k^{\text{eff}}\hat{\sigma}_k^z\right)+g_{\text{eff}}(\hat{\sigma}_1^+\hat{\sigma}_2^-+H.C.)\\
        \omega_k^{\text{eff}}&=\omega_k-g_{k\text{C}}^2\left(\frac{1}{\Delta_k}+\frac{1}{\Sigma_k}\right),\ k\in\{1,2\}
    \end{align}
\end{subequations}
where $g_{\text{eff}}$ is the effective coupling strength between qubits, which is determined by the expression in Eq.(\ref{equ:1}). If $\Delta_k=\omega_\text{C}-\omega_k>0$, i.e., when the frequency of the tunable coupler is above both frequencies of qubits, the virtual exchange interaction term $g_{1\text{C}}g_{2\text{C}}/\Delta_k>0$, and $g_{\text{eff}}$ can be tuned from negative to positive monotonically by increasing the frequency of the tunable coupler. Therefore, a critical value $\omega^{\text{off}}_\text{C}$ can always be reached to turn off the effective qubit-qubit coupling, i.e., $g_{\text{eff}}(\omega^{\text{off}}_\text{C})=0$. It is important to note that we assume that the qubits are far enough apart spatially so that the direct qubit-qubit capacitance $C_{12}$ is negligible. As a result, $g_{\text{eff}}$ does not depend on $C_{12}$ and is replaced by the direct capacitances between qubits and connecting pads $C_{1\text{P}}$, $C_{2\text{P}}$ and the direct capacitance between connecting pads $C_{\text{P}12}$ as derived in Appendix \ref{sec:A}.

With the TCCP architectures, both grounded and floating transmons can be used as the qubits and the tunable couplers. Previous paper \cite{FloatingCoupler} proposes a tunable coupler that achieves a vanishing effective qubit-qubit coupling $g_{\text{eff}}=0$ without requiring direct capacitive coupling between the qubits. However, the design of their tunable coupler is only compatible with floating transmon, which has two superconducting pads connected by a SQUID. To incorporate readout resonators, airbridges, Purcell filters, TSV, and other forms of the three-dimensional integration to scale up, the pads of the floating tunable couplers and even the floating qubits need to be elongated significantly to provide enough wiring space. This elongation degrades the performance of the couplers and qubits, such as coherence time $T_1$ and fidelity, because the elongated pads increase the participation ratio of some lossy interfaces, such as the metal-air (MA), metal-substrate (MS), and substrate-air (SA) interfaces \cite{T1SurfaceLoss,Kelly,SurfaceLoss}. In contrast, using grounded transmon with a relatively small physical size as the tunable couplers and qubits with the TCCP architectures can reduce these problems. Furthermore, it can also more effectively reduce the parasitic capacitances caused by other layer along the stacking direction and improve the convenience and flexibility of layout design in the three-dimensional integration.

In order to avoid additional dissipation introduced by the TCCP, the size of the connecting pads should much small than the microwave wavelength. For superconducting qubits, their operating frequencies are typically lower than 10GHz, corresponding to a wavelength of 9mm in silicon substrate \cite{Microwave,Resonators}. Thus, to maintain the lumped nature of the connecting pads, their size should be far less than the corresponding wavelength. Consequently, the wiring distance between the qubits can reach millimeter scale, which is appropriate for large-scale integration of qubits.

\section{Other Architectures of TCCP}

\begin{figure}[htb]
  \center{\includegraphics[width=8cm]  {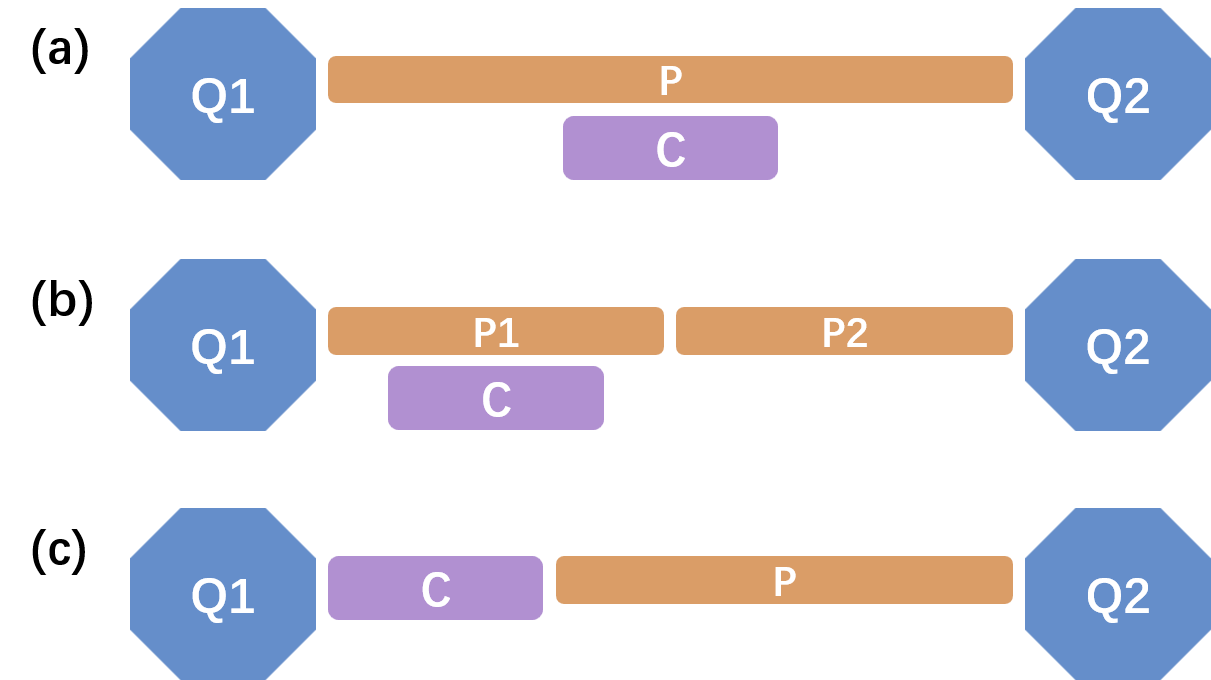}}
  \caption{The diagram of other different TCCP architectures. (a) Only use one connecting pad. (b) Use two connecting pads while the tunable coupler is in an asymmetrical position. (c) Use one connecting pad while the tunable coupler is in an asymmetrical position.}
  \label{fig:5}
\end{figure}

By introducing the connecting pads, we can achieve other different coupling architectures, as shown in FIG.\ref{fig:5}, in addition to the architecture shown in FIG.\ref{fig:1}(c). FIG.\ref{fig:5}(a) shows an architecture that uses only one connecting pad. Its analytic calculations are provided in Appendix \ref{sec:A}. Additionally, we can place the tunable coupler in an asymmetrical position with respect to the two qubits and the connecting pads. For example, in the case with two connecting pads, we can capacitively couple the tunable coupler to only one of the connecting pads, as shown in FIG.\ref{fig:5}(b). In the case with one connecting pad, we can sandwich the tunable coupler between the connecting pad and one of the qubits, as shown in FIG.\ref{fig:5}(c). These different architectures can easily achieve their own tunable coupling parameters, expanding our design possibilities.

\section{Designs of TCCP Architectures}

\begin{figure}[htb]
  \center{\includegraphics[width=8.5cm]  {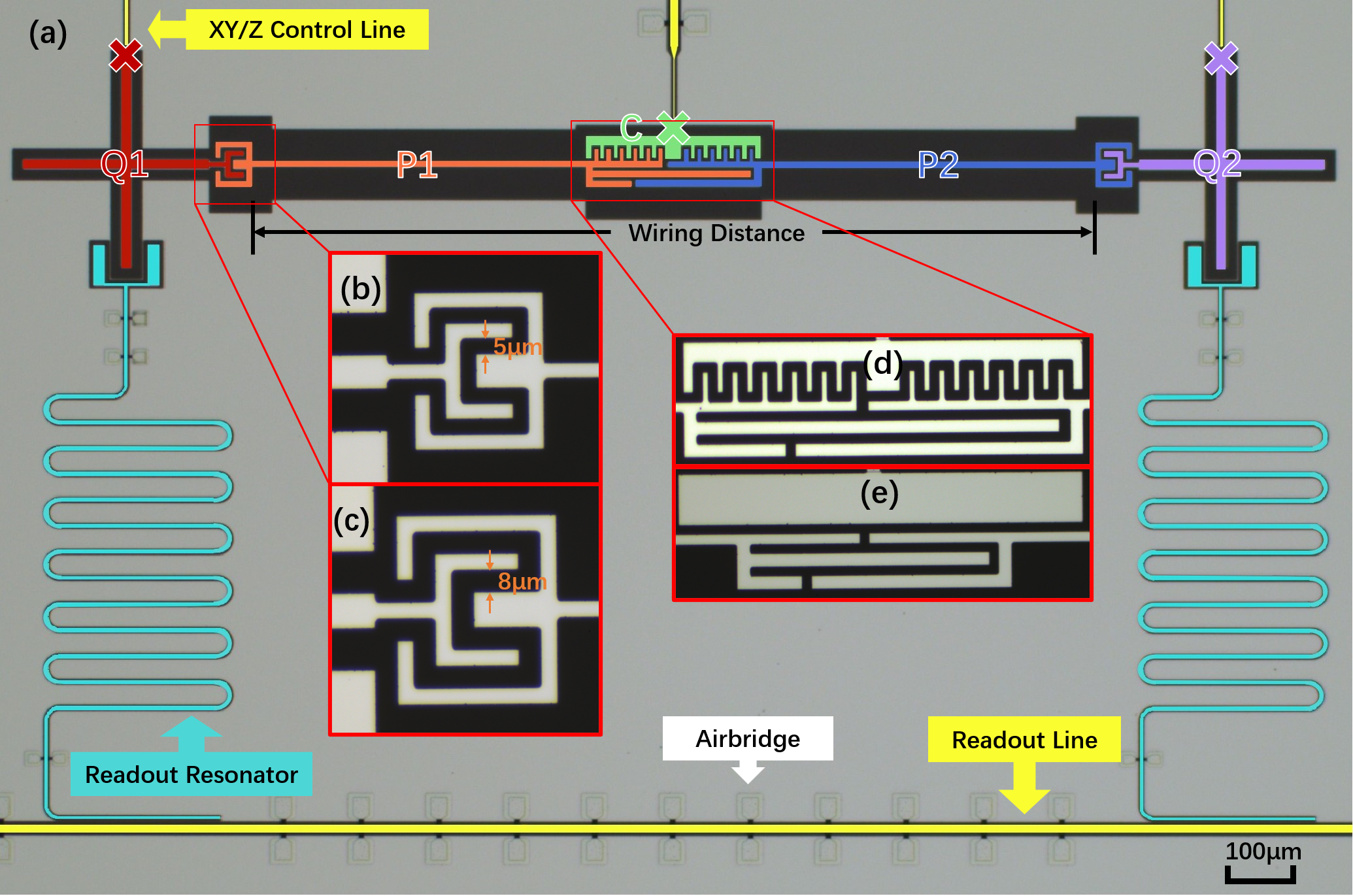}}
  \caption{(a) The optical microscope of the TCCP architecture Design A. The five pads of two qubits, two connecting pads, and one coupler are labeled with different colors and designated as Q1, Q2, P1, P2 and C. The Josephson junctions are marked by crosses on the qubits and the tunable coupler. Additionally, the image indicates the XY/Z control lines, readout resonators, readout line, and airbridges. (b-c) The coupling areas between the qubit and the connecting pad have either 5$\mu$m or 8$\mu$m etching gap. (d-e) The coupling areas between the tunable coupler and the connecting pad have either interdigital or planiform shape.}
  \label{fig:2}
\end{figure}

To validate the feasibility of the TCCP architectures, we designed and fabricated a device with four distinct TCCP architectures (Design A, B, C and D), each of which contained two or one connecting pads based on the lumped-element circuit model shown in FIG.\ref{fig:A1}(a) or (c). The optical micrograph of Design A is shown in FIG.\ref{fig:2}(a), and some detailed design, simulation and characterized parameters are listed in Table \ref{table:1}. To simplify the analysis, we assume that all ground capacitances of the qubits are identical ($C_{\text{Q1G}}=C_{\text{Q2G}}=C_{\text{QG}}$), as are the ground capacitances of the connecting pads ($C_{\text{P1G}}=C_{\text{P2G}}=C_{\text{PG}}$), the coupling capacitances between the qubits and the connecting pads ($C_{\text{1P}}=C_{\text{2P}}=C_{\text{QP}}$), and the coupling capacitances between the tunable coupler and the connecting pads ($C_{\text{P1C}}=C_{\text{P2C}}=C_{\text{PC}}$). In addition, $C_{\text{CG}}$ represents the ground capacitance of the tunable coupler, while $C_{\text{P12}}$ represents the coupling capacitance between two connecting pads (the designs with one connecting pad lack such capacitance). By carefully designing, simulating, and calculating these capacitances, we ensured that $g_{\text{eff}}$ of these four designs had a wide enough tunable range and could be turned off.

In the TCCP architectures, the capacitances $C_{\text{QP}}$ and $C_{\text{PC}}$ should be relatively large to turn off, i.e., $g_{\text{eff}}(\omega_\text{C}^\text{off})=0$, and achieve a wide enough tunable range (see Appendix \ref{sec:A} for details). In order to achieve the desired value of $C_{\text{QP}}$, we increased the coupling area as shown in FIG.\ref{fig:2}(b). But increasing the coupling area is limited by the area of the qubit. So we then reduced the etching gap as shown in FIG.\ref{fig:2}(c). However, reducing the etching gap is limited by the accuracy of the photolithography instrument. It would also introduce dense impurities which could degrade the performance of qubits. Therefore, we varied the coupling areas and etching gaps simultaneously to obtain an appropriate $C_{\text{QP}}$ without the degradation. On the other hand, due to the limitation of our readout bandwidth, each readout line can read up to 10 qubits. Thus, we designed two different wiring distances as shown in FIG.\ref{fig:2}(a), 1280$\mu$m and 820$\mu$m, which correspond to 10 qubits and 6 qubits per readout line, respectively. We designed four groups of parameters as shown in Table \ref{table:1}, and validated these designs in subsequent characterizations to assess if they degraded the performance of qubits. From the results, it appears that these designs worked well.

\begin{table*}[htb]
  \begin{tabular}{lllllll}
      \hline\hline
      \multicolumn{2}{l}{Design}                                             & \multicolumn{1}{l}{A}            & B     &    & \multicolumn{1}{l}{C}         & D            \\ \hline
      \multicolumn{2}{l}{Architecture}                                       & \multicolumn{2}{l}{two connecting pads$\ $}  & $\ \ $       & \multicolumn{2}{l}{one connecting pad}           \\ \cline{3-4}\cline{6-7}
      \multicolumn{2}{l}{Wiring Distance ($\mu$m)}                               & \multicolumn{1}{l}{1280$\ \ \ \ \ \ \ \ \ $}         & 820   &    & \multicolumn{1}{l}{1280$\ \ \ \ \ \ \ \ \ $}      & 820          \\ 
      \multicolumn{2}{l}{Etching Gap of Coupling ($\mu$m)}          & \multicolumn{1}{l}{5}            & 8    &     & \multicolumn{1}{l}{8}         & 5            \\ 
      \multicolumn{1}{l}{Ground Capacitance $C_{\text{QG}}$ (fF)} & & \multicolumn{1}{l}{72.5}         & 72.4   &   & \multicolumn{1}{l}{71.7}      & 71.8         \\ 
      \multicolumn{1}{l}{Ground Capacitance $C_{\text{CG}}$ (fF)}                                            &   & \multicolumn{1}{l}{25.1}         & 32.5   &   & \multicolumn{1}{l}{36}        & 28.2         \\ 
      \multicolumn{1}{l}{Ground Capacitance $C_{\text{PG}}$ (fF)}                                            &   & \multicolumn{1}{l}{61.7}         & 42.2  &    & \multicolumn{1}{l}{108.8}     & 74.7         \\
      \multicolumn{1}{l}{Coupling Capacitance $C_{\text{QP}}$ (fF)} &   & \multicolumn{1}{l}{11.5}         & 11    &    & \multicolumn{1}{l}{6.9}       & 8.8          \\ 
      \multicolumn{1}{l}{Coupling Capacitance $C_{\text{PC}}$ (fF)}                                            &   & \multicolumn{1}{l}{17.8}         & 12.3 &     & \multicolumn{1}{l}{23.7}      & 32.8         \\ 
      \multicolumn{1}{l}{Coupling Capacitance $C_{\text{P12}}$ (fF)}                                            &  & \multicolumn{1}{l}{21}           & 13.6   &   & \multicolumn{1}{l}{-}         & -            \\ 
      \multicolumn{2}{l}{Equivalent Ground Capacitance of Qubit (fF)}    & \multicolumn{1}{l}{91.6}         & 90.7    &  & \multicolumn{1}{l}{87.2}      & 88.8         \\ 
      \multicolumn{2}{l}{Equivalent Ground Capacitance of Coupler (fF)}   & \multicolumn{1}{l}{60.5}         & 59.3   &   & \multicolumn{1}{l}{62.8}      & 59.3         \\ 
      \multicolumn{2}{l}{Anharmonicity of Qubit $\alpha_{\text{Q}}$ (MHz)}                 & \multicolumn{1}{l}{232}          & 234   &    & \multicolumn{1}{l}{245}       & 240          \\ 
      \multicolumn{2}{l}{Anharmonicity of Coupler $\alpha_{\text{C}}$ (MHz)}                & \multicolumn{1}{l}{361}          & 369  &     & \multicolumn{1}{l}{346}       & 368          \\ 
      \multicolumn{2}{l}{Maximum Frequency of Qubit $\omega_{\text{Q}}$ (GHz)}            & \multicolumn{1}{l}{5.59}         & 5.62   &   & \multicolumn{1}{l}{5.73}      & 5.67         \\ 
      \multicolumn{2}{l}{Maximum Frequency of Coupler $\omega_{\text{C}}$ (GHz)}            & \multicolumn{1}{l}{6.83}         & 6.89  &    & \multicolumn{1}{l}{6.7}       & 6.89         \\ 
      \multicolumn{2}{l}{Qubit-Qubit Coupling Strength $g_{12}$ (MHz)}            & \multicolumn{1}{l}{9.62}         & 10.94 &    & \multicolumn{1}{l}{11.22}     & 22.8         \\ 
      \multicolumn{2}{l}{Qubit-Coupler Coupling Strength $g_{\text{QC}}$ (MHz)}          & \multicolumn{1}{l}{83.5}         & 79   &     & \multicolumn{1}{l}{43}        & 91           \\ 
      \multicolumn{2}{l}{Effective Coupling Strength $g_{\text{eff}}$ (MHz)}     & \multicolumn{1}{l}{1.91}         & 4.18    &  & \multicolumn{1}{l}{8.78}      & 13.53        \\ 
      \multicolumn{2}{l}{Average Decoherence Time $\overline{T}_1$ ($\mu$s)}           & \multicolumn{1}{l}{12.1}         & 23.4   &   & \multicolumn{1}{l}{23.9}        & 24.0           \\ 
      \multicolumn{2}{l}{Dephasing Time $T_2$ ($\mu$s)}                     & \multicolumn{1}{l}{7.2}          & 8.6  &     & \multicolumn{1}{l}{10.5}       & 12.2         \\ \hline\hline
      \end{tabular}
  \center\caption{Some detailed design, simulation and characterized parameters of four designs of the connecting architectures, where only $\overline{T}_1$ and $T_2$ are the result of measurement. In addition, $\overline{T}_1$ and $T_2$ of the uncoupled qubits as the control group are 23.0$\mu$s and 12.5$\mu$s, respectively.} 
  \label{table:1}
\end{table*}

\section{Sample Fabrication and Measurement}
We have fabricated a chip that contained four pairs of transmon qubits coupled using the TCCP with different design parameters. A few uncoupled transmon qubits were also fabricated on the chip as the control group. The chip also contained readout resonators, XY/Z control lines, and airbridges. The fabrication process is similar to those described previously \cite{Zhaoshoukuan,Guoxueyi}. After packaging the sample, it was cooled to approximately 10mK in a dilution refrigerator for measurement \cite{Dispersive}. Details of the measurement setup can be found in Appendix \ref{sec:C}.

The measurement showed that the qubits coupled by TCCP showed comparable performance to those of the control group as shown in Table \ref{table:1}. To evaluate the performance of the qubits, we measured the energy relaxation time $T_1$ at various frequencies and obtained the averaged $T_1$ (denoted as $\overline{T}_1$). To achieve a comprehensive evaluation, we averaged $T_1$ over the frequency range from the maximum frequency of the qubit to 1GHz below it. We found that $\overline{T}_1$ of the TCCP architectures was comparable with that of the qubit in the control group, 23.0$\mu$s. Additionally, we measured the dephasing time $T_2$ and obtained the pure dephasing time $T_\phi$ ($1/T_2=1/2T_1+1/T_\phi$) around the sweet point of the qubits. All the qubits showed comparable performance on $T_{\phi}$, with the maximum $T_{\phi}$ values of the uncoupled qubits to be 12.5$\mu$s. Among them, Design A exhibited a lower $\overline{T}_1$, which might be attributed to the imperfections left during the fabrication process or design parameters.

\begin{figure}[htb]
  \center{\includegraphics[width=8cm]  {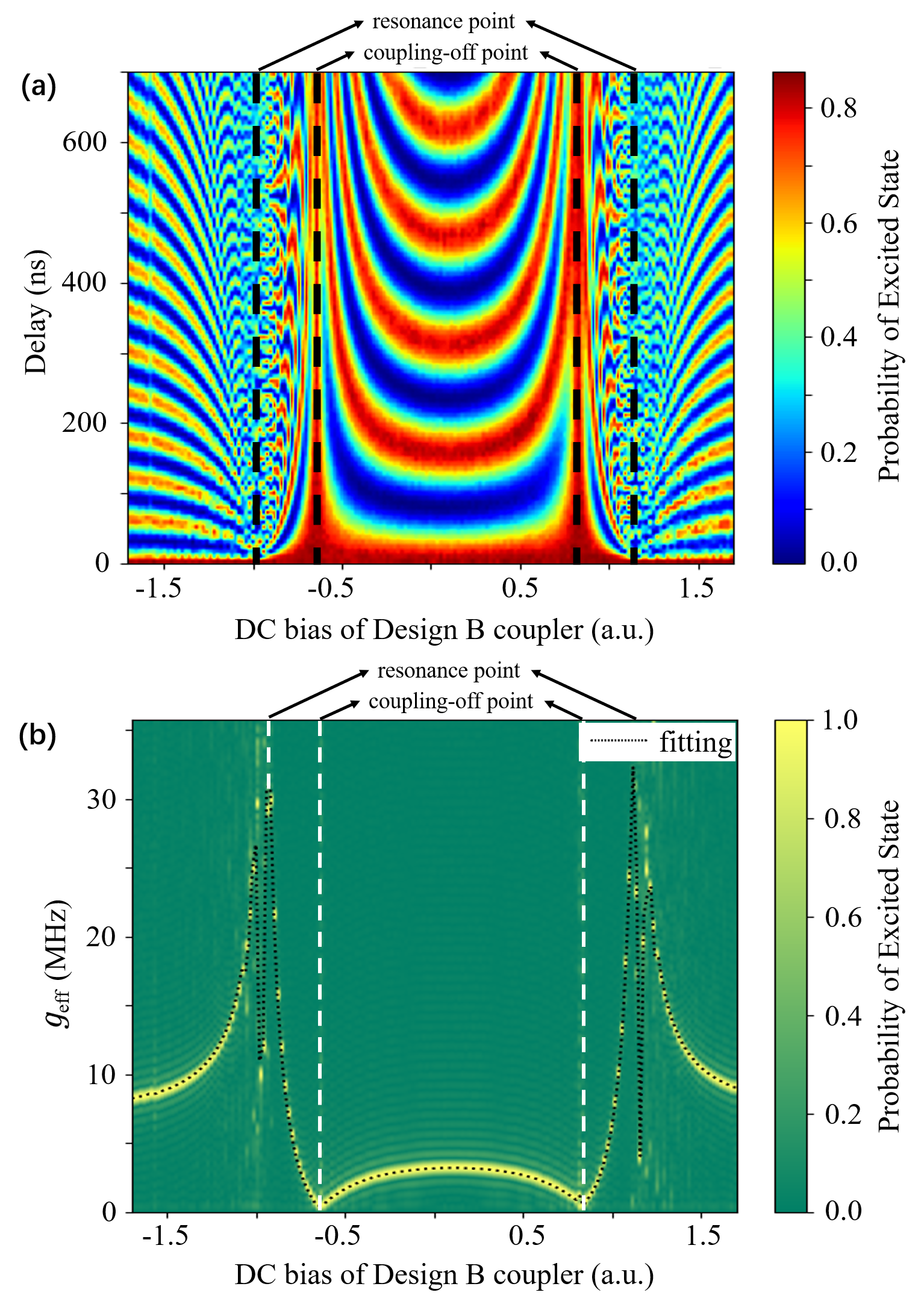}}
  \caption{(a) The spectral diagram of the SWAP operation performed on Design B for different frequencies of the tunable coupler, during a period of delay. The abscissa represents the dimensionless parameter of the DC bias applied to the tunable coupler, which correlates inversely with the frequency of the tunable coupler. The colors indicate the probability that one qubit is in the excited state. (b) $g_{\text{eff}}$ as a function of the frequency of the tunable coupler was obtained by Fourier transforming the data in (a). The dashed line represents the fitting result.}
  \label{fig:3}
\end{figure}

To investigate the coupling characteristics, we first measured the frequency of the tunable coupler under different DC bias, and obtained the direct coupling strength between the qubit and the tunable coupler $g_{\text{QC}}$. We then demonstrated a SWAP gate between two qubits coupled by the TCCP to obtain $g_{\text{eff}}$. In this study, we focused on Design B, and presented the results of $g_{\text{eff}}$ as the function of DC bias applying to the coupler in FIG.\ref{fig:3} (refer to Appendix \ref{sec:E} for more details on the measurements related to TCCP). The spectral diagram is shown in FIG.\ref{fig:3}(a), and the $g_{\text{eff}}$ was obtained by performing Fourier transform on it as shown in FIG.\ref{fig:3}(b). It is important to note that the frequency of the tunable coupler should not be too close to the frequencies of the qubits, as this can lead to a significant loss of energy transferred by the qubits and make the TCCP architectures work poorly. We therefore assessed the available range of $g_{\text{eff}}$ by observing the attenuation of the energy exchange of the qubits in the spectral diagram. When an obvious attenuation occurred, we concluded that the corresponding $g_{\text{eff}}$ was unavailable \cite{Sunluyan}. We found that $g_{\text{eff}}$ of Design B could be modulated between $+$3MHz and $-$25MHz, which is a sufficiently wide tunable range to enable most experimental schemes.

The use of one or two connecting pads can facilitate achieving different designs for the tunable range of $g_{\text{eff}}$. Compared to the architecture with one connecting pad, the architecture with two connecting pads has an additional capacitance $C_{\text{P12}}$ between the connecting pads, which provides more flexible adjustment of $g_{\text{eff}}$ when the tunable coupler is at the sweet point. However, this also requires larger $C_{\text{QP}}$ and $C_{\text{PC}}$ in the architecture with two connecting pads to achieve the desired tunable range of $g_{\text{eff}}$. Moreover, based on both numerical simulation and experimental observations, we found that the residual ZZ interaction was higher in the architecture with one connecting pad. However, the residual ZZ interactions of both architectures were less than 1MHz in a wide range of the coupler frequency as shown in Appendix \ref{sec:F}.

\begin{figure}[htb]
  \center{\includegraphics[width=8cm]  {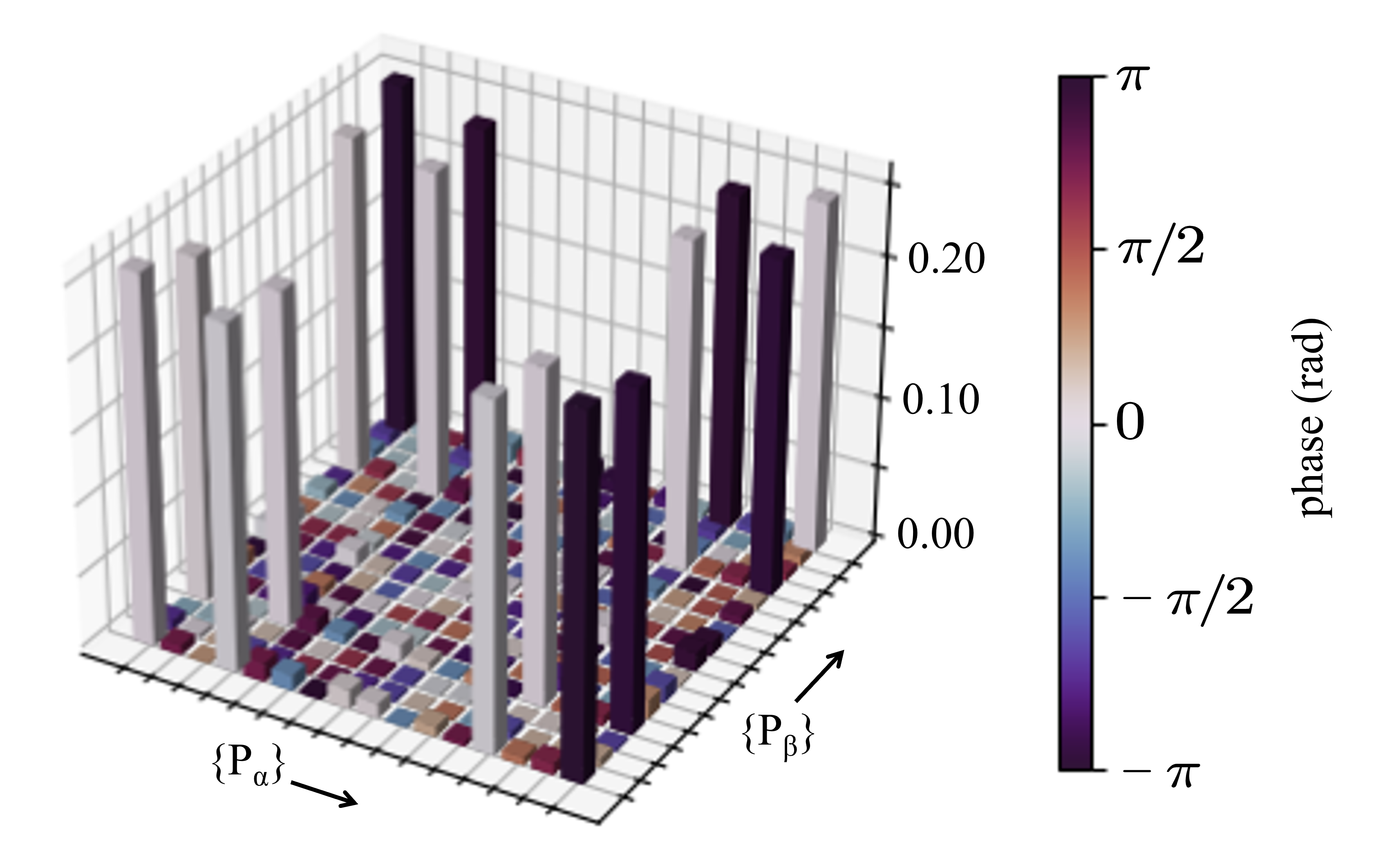}}
  \caption{The $\chi$ matrix derived from Quantum Processing Tomography (QPT) of the adiabatic CZ gate of Design D. In the direction of the arrow, the Pauli operators of $\{P_\alpha\}$ and $\{P_\beta\}$ in turn are II, IX, IY, IZ, XI, XX, XY, XZ, YI, YX, YY, YZ, ZI, ZX, ZY, ZZ.}
  \label{fig:4}
\end{figure}

We have demonstrated the practical performance of the adiabatic CZ gate on the TCCP architectures. To achieve this, we applied a bias current to the tunable coupler, which caused its frequency to gradually decrease and rise. The adiabatic process allowed the accumulation of the relative phase between the qubits through the tunable coupler. We presented an example of the adiabatic CZ gate with Design D in FIG.\ref{fig:4}, where two qubits and one tunable coupler were positioned at 5.11GHz, 5.64GHz, and 6.404GHz, respectively. To reduce the residual ZZ interaction between the qubits and mitigate the errors, we adjusted the initial frequency of the tunable coupler to 6.404GHz instead of the sweet point, which was around 7.1GHz. Accurately preparing the initial state is crucial to ensuring the dependability of Quantum Processing Tomography (QPT), which is used to measure the fidelity of the adiabatic CZ gate. To prevent non-adiabatic leakage, we designed a smooth trajectory for the movement of the frequency of the tunable coupler. The entire process adiabatically accumulated the relative phase $\pi$ on the $|101\rangle$ state. We did QPT under the $\chi$ matrix representation \cite{Nielsen}:
\begin{gather}
  \Lambda(\rho)=\sum_{\alpha\beta}\chi_{\alpha\beta}P_\alpha\rho P_\beta\\
  \chi_{\alpha\beta}=\langle\text{CZ}\otimes\text{CZ}|P_\alpha\cdot P_\beta\rangle
\end{gather}
where $\Lambda(\rho)$ is the quantum operation that experimentally realizes the adiabatic CZ gate, $\rho$ is the density matrix of the qubits, and $\{P_\alpha\}$ and $\{P_\beta\}$ are the complete Pauli basis of the space of the two qubits. After optimization, the fidelity of the adiabatic CZ gate of Design D reached $96.2\%$, and that of Design B measured by the same method was $95.6\%$. These results demonstrated the practicality of the TCCP architectures for the superconducting qubits with the tunable coupling.

\section{Flip-Chip Process with TCCP Architectures}

\begin{figure}[htb]
  \center{\includegraphics[width=8.5cm]  {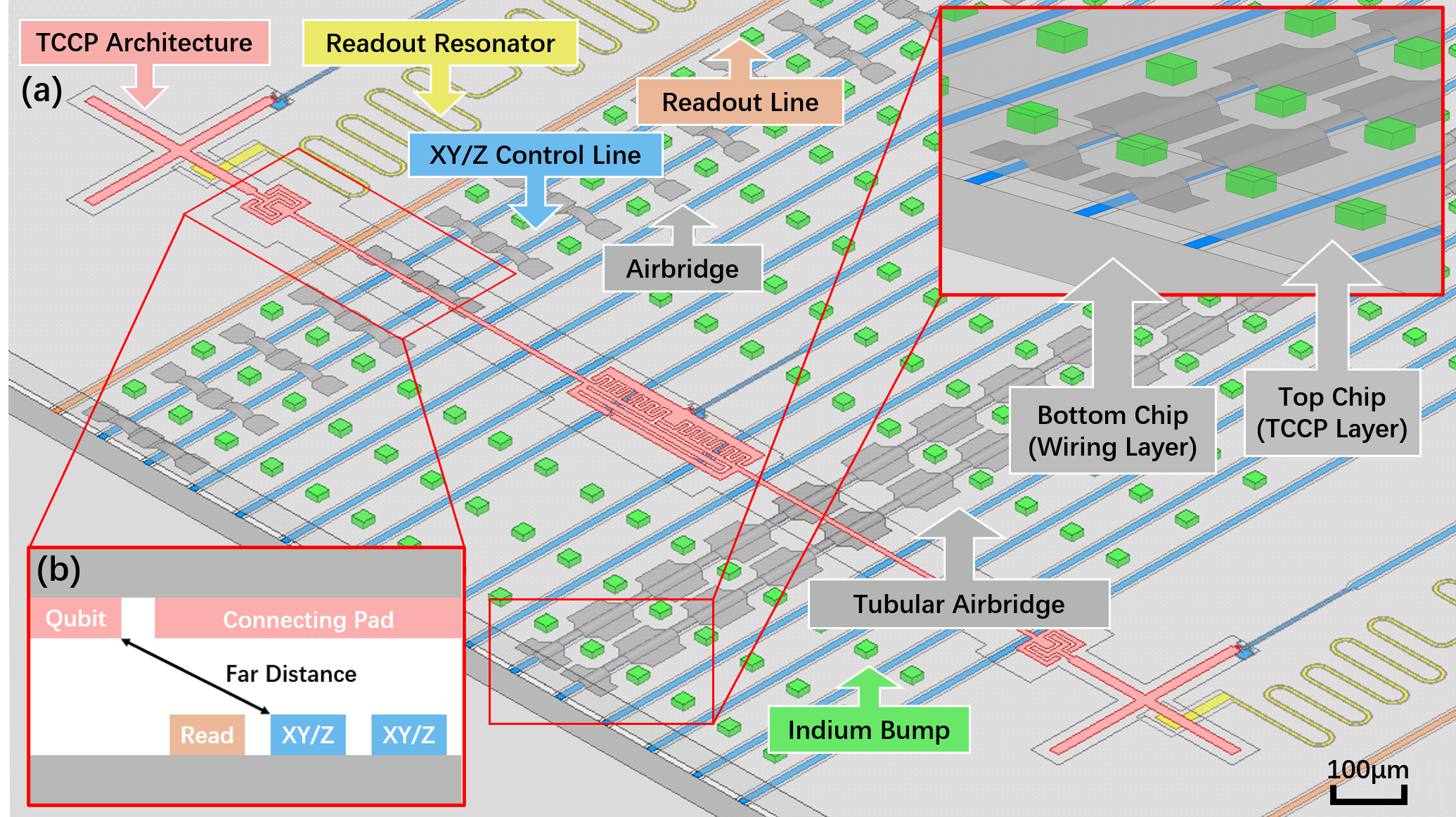}}
  \caption{(a) A schematic diagram of the flip-chip process using the TCCP architecture. The TCCP architecture on the top chip provides sufficient space for the bottom chip to accommodate readout resonators, readout lines, control lines, (tubular) airbridges, indium bumps, and other components. (b) Schematic cross-section of the marked location in (a). Because of the connecting pads, the qubit is far away from the control lines.}
  \label{fig:2-1}
\end{figure}

The schematic diagram FIG.\ref{fig:2-1}(a) demonstrates the feasibility of the flip-chip design \cite{FlipReview,Flipchip1,Flipchip2} using the TCCP architectures. Sufficient space is available between two qubits to accommodate readout resonators, readout lines, control lines, (tubular) airbridges, indium bumps, and other components. Due to the introduction of the connecting pads, most control lines driving other qubits can be placed far from the qubit as shown in FIG.\ref{fig:2-1}(b), thus greatly reducing the crosstalk between them. Simultaneously, for the control lines in close proximity to the qubit, rows of indium can be used instead of indium bumps to block the crosstalk more effectively. Furthermore, the relatively small size of the qubits and the tunable couplers can reduce the parasitic capacitance caused by other layer along the stacking direction effectively. By adjusting each pad, we are able to achieve a wide enough tunable range and turn-off point for $g_{\text{eff}}$ in the flip-chip design, which is comparable to those in the plane design.

Because of the introduction of the connecting pads, the TCCP architectures can also be utilized for connecting qubits between different chips, such as pressing two top chips (qubits layer) to one bottom chip (wiring layer). The TCCP architectures enable the cross-chip tunable coupling of the adjacent qubits between the two top chips. Taking FIG.\ref{fig:5}(c) as an example, we can prepare Q1 and C on a top chip, and Q2 on another top chip, and then perform the cross-chip coupling through P1. We can also directly prepare the whole P on the bottom chip and couple it to C and Q2 respectively. Or, we can prepare part of P on the top chip coupled to Q2 and C and the other part of P on the bottom chip for the cross-chip tunable coupling, and connect them with indium bumps. Of course, other architectures can be used to achieve different ways of the cross-chip tunable coupling as well.

\section{Conclusion}
We have introduced the connecting pads between the qubits and the tunable coupler to design a new tunable coupler called TCCP architectures. By eliminating the direct capacitive coupling between the qubits, the TCCP architectures can separate the two qubits further apart, obtaining sufficient wiring space for the flip-chip process, reducing the crosstalk from other control lines to qubits, and providing a new scheme for the modularization of the superconducting quantum processors. We have measured the performance of our samples prepared using the TCCP architectures with various designs and verified that the TCCP architectures have well working performance. However, we believe that our designs and measurements are not optimal, leaving room for the further development of the TCCP architectures. In conclusion, the TCCP architectures offer a promising approach for the realization of the large-scale superconducting quantum processors.

\begin{acknowledgments}
  We thank Zheng-Hang Sun for helpful discussions. This work was supported by the Synergetic Extreme Condition User Facility (SECUF). Devices were made at the Nanofabrication Facilities at Institute of Physics, CAS in Beijing. This work was supported by: the National Natural Science Foundation of China (Grants No. 12204528, 92265207, T2121001, 11934018, 12005155, 11904393, 92065114, 12204528, 11875220, and 12047502), Key Area Research and Development Program of Guangdong Province, China (Grants No. 2020B0303030001, 2018B030326001), Beijing Natural Science Foundation (Grant No. Z200009), Innovation Program for Quantum Science and Technology (Grant No. 2021ZD0301800), Strategic Priority Research Program of Chinese Academy of Sciences (Grant No. XDB28000000), and Scientific Instrument Developing Project of Chinese Academy of Sciences (Grant No. YJKYYQ20200041).
\end{acknowledgments}

\appendix

\renewcommand{\thefigure}{\Alph{section}\arabic{figure}}
\renewcommand{\thetable}{\Alph{section}\arabic{table}}

\section{Circuit Principle of TCCP Architectures\label{sec:A}}
\setcounter{figure}{0}
\setcounter{table}{0}

\begin{figure*}[htb]
  \center{\includegraphics[width=17cm]  {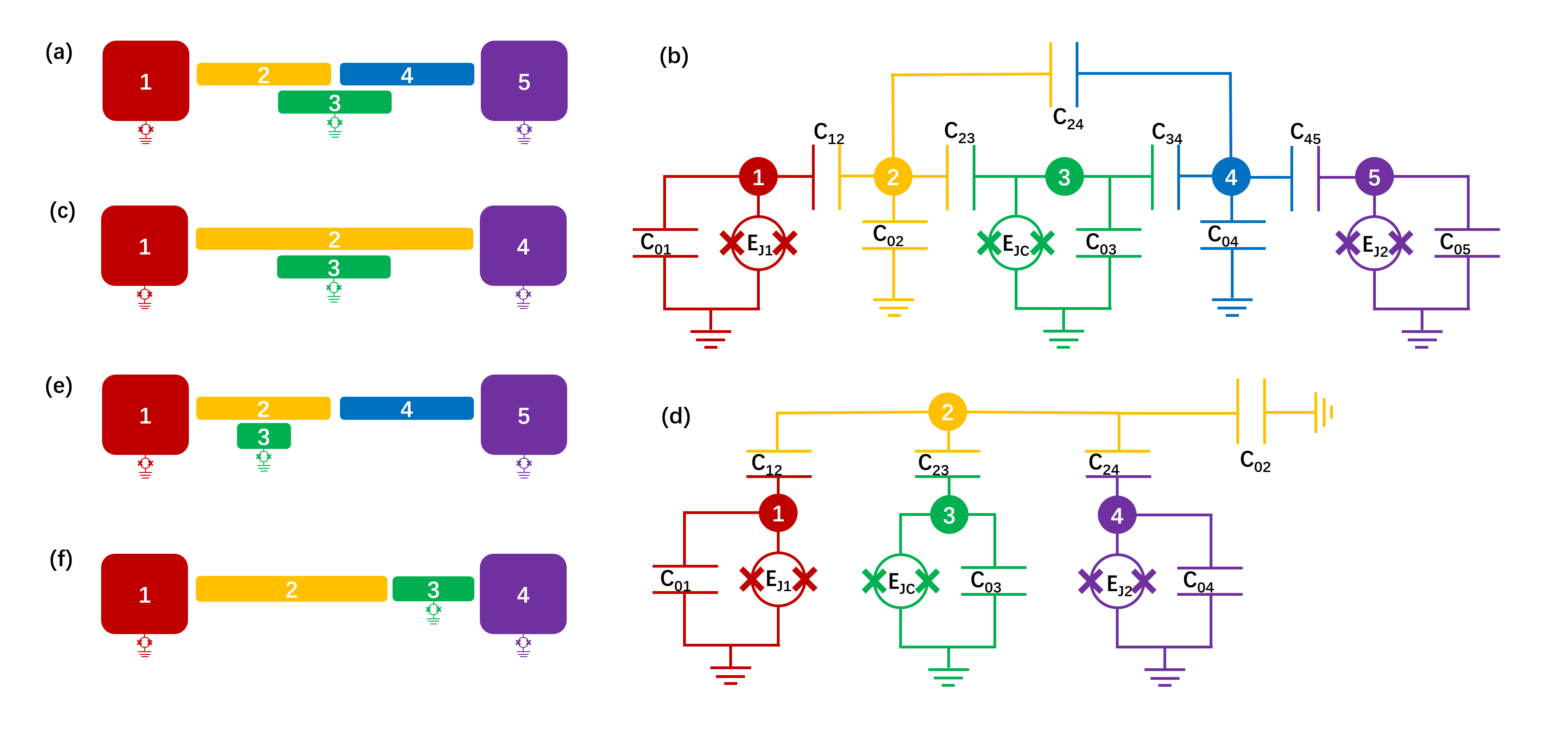}}
  \caption{(a) The electrode schematic diagram of the TCCP architecture with two connecting pads. (b) The equivalent lumped-element circuit model of (a). (c) The electrode schematic diagram of the TCCP architecture with one connecting pad. (d) The equivalent lumped-element circuit model of (c). (e) The electrode schematic diagram of the TCCP architecture with two connecting pads and the asymmetrically placed tunable coupler. (f) The electrode schematic diagram of the TCCP architecture with one connecting pad and the asymmetrically placed tunable coupler.}
  \label{fig:A1}
\end{figure*}

We can directly write the Lagrangian of the TCCP architecture with two connecting pads shown in FIG.\ref{fig:A1}(a-b):
\begin{subequations}
  \begin{align}
      \mathcal{L}=&T-U\\
      T=&\sum_{j=1}^5\frac12C_{0j}\dot{\Phi}_j^2+\sum_{j,k\neq0}\frac12C_{jk}(\dot{\Phi}_j-\dot{\Phi}_k)^2\\
      U=&-E_{J1}(\phi_{\text{e}1})\cos(\phi_1+\phi_{01})-E_{J\text{C}}(\phi_{\text{e}\text{C}})\cos(\phi_3+\phi_{0\text{C}})\notag\\
      &-E_{J2}(\phi_{\text{e}2})\cos(\phi_5+\phi_{02})
  \end{align}
\end{subequations}
where $C_{jk}$ includes only the capacitances present in FIG.\ref{fig:A1}(b), excluding the ground capacitances. $E_{Jl}(\phi_{\text{e}l})$ and $\phi_{0l},\ l\in\{1,\text{C},2\}$ can be directly obtained from the asymmetrical DC-SQUID conclusion \cite{Asymmetric}:
\begin{align}
  E_{Jl}(\phi_{\text{e}l})=&\sqrt{E_{J\text{s}l}^2+E_{J\text{b}l}^2+2E_{J\text{s}l}E_{J\text{b}l}\cos(\phi_{\text{e}l})}\label{equ:A1}\\
  \phi_{0l}=&\tan^{-1}\left[\frac{E_{J\text{s}l}-E_{J\text{b}l}}{E_{J\text{s}l}+E_{J\text{b}l}}\tan\left(\frac{\phi_{\text{e}l}}{2}\right)\right]
\end{align}
The maximum value of Eq.(\ref{equ:A1}) is $E_{Jl}^{\text{max}}=E_{J\text{b}l}+E_{J\text{s}l}$, where $E_{J\text{b}l}$ and $E_{J\text{s}l}$ are the larger and smaller Josephson energy in SQUID $l$ respectively. The flux phase at node $j$ is $\phi_j=2\pi\Phi_j/\Phi_0,\ j\in\{1,\ldots,5\}$ with $\Phi_0=h/2e$ to be the flux quantum and $\Phi_j$ to be the magnetic flux at node $j$. $\phi_{\text{e}l}=2\pi\Phi_{\text{e}l}/\Phi_0$ is the external flux bias through the SQUID $l$. To simplify the calculations, we will replace $E_{Jl}(\phi_{\text{e}l})$ with $E_{Jl}$ in the following. The flux at node $j$ is defined by its voltage, $\Phi_j=\int_{-\infty}^tdt'V_j(t')$, where $V_j$ is the voltage at node $j$. The relationship between the canonical momentum $Q_j$ and the coordinate $\Phi_j$ is given by $Q_j=\partial\mathcal{L}/\partial\dot{\Phi}_j$, which can be written in matrix form as $\bm{Q}=\bm{C}\dot{\Phi}$. The calculation for each canonical momentum is:
\begin{subequations}
  \begin{align}
      Q_1=&C_{01}\dot{\Phi}_1+C_{12}(\dot{\Phi}_1-\dot{\Phi}_2)\\
      Q_2=&C_{02}\dot{\Phi}_2+C_{12}(\dot{\Phi}_2-\dot{\Phi}_1)+C_{23}(\dot{\Phi}_2-\dot{\Phi}_3)\notag\\
      &+C_{24}(\dot{\Phi}_2-\dot{\Phi}_4)\\
      Q_3=&C_{03}\dot{\Phi}_3+C_{23}(\dot{\Phi}_3-\dot{\Phi}_2)+C_{34}(\dot{\Phi}_3-\dot{\Phi}_4)\\
      Q_4=&C_{04}\dot{\Phi}_4+C_{24}(\dot{\Phi}_4-\dot{\Phi}_2)+C_{34}(\dot{\Phi}_4-\dot{\Phi}_3)\notag\\
      &+C_{45}(\dot{\Phi}_4-\dot{\Phi}_5)\\
      Q_5=&C_{05}\dot{\Phi}_5+C_{45}(\dot{\Phi}_5-\dot{\Phi}_4)
  \end{align}
\end{subequations}
The final matrix form is:
\begin{equation}
  \left(\begin{matrix}
      Q_1 \\ Q_2 \\ Q_3 \\ Q_4 \\ Q_5
  \end{matrix}\right)=\left(\begin{matrix}
      C_{1S} & -C_{12} & 0 & 0 & 0 \\
      -C_{12} & C_{2S} & -C_{23} & -C_{24} & 0 \\
      0 & -C_{23} & C_{3S} & -C_{34} & 0 \\
      0 & -C_{24} & -C_{34} & C_{4S} & -C_{45} \\
      0 & 0 & 0 & -C_{45} & C_{5S}
  \end{matrix}\right)\left(\begin{matrix}
      \dot{\Phi}_1 \\ \dot{\Phi}_2 \\ \dot{\Phi}_3 \\ \dot{\Phi}_4 \\ \dot{\Phi}_5
  \end{matrix}\right)\label{equ:A8}
\end{equation}
where $C_{1S}=C_{01}+C_{12},\ C_{2S}=C_{02}+C_{12}+C_{23}+C_{24},\ C_{3S}=C_{03}+C_{23}+C_{34},\ C_{4S}=C_{04}+C_{24}+C_{34}+C_{45},\ C_{5S}=C_{05}+C_{45}$. The Hamiltonian of the system is given by $H=\frac12\bm{Q}^T\bm{C}^{-1}\bm{Q}+U$. As shown by inverting the capacitance matrix in Eq.(\ref{equ:A8}), we can ignore the free particle variables without corresponding potential energy, namely $\Phi_2$ and $\Phi_4$. This allows us to represent the matrix form with the following notations:
\begin{equation}
  \left(\begin{matrix}
      \dot{\Phi}_1 \\ \dot{\Phi}_3 \\ \dot{\Phi}_5
  \end{matrix}\right)=\left(\begin{matrix}
      A_{11} & A_{13} & A_{15} \\
      A_{31} & A_{33} & A_{35} \\
      A_{51} & A_{53} & A_{55} 
  \end{matrix}\right)\left(\begin{matrix}
      Q_1 \\ Q_3 \\ Q_5
  \end{matrix}\right)
\end{equation}
If we assume that the TCCP architecture is symmetric, with $C_{12}=C_{45},\ C_{23}=C_{34},\ C_{02}=C_{04},\ C_{01}=C_{05}$, we can simplify the expression of the notations to:
\begin{widetext}
\begin{subequations}
  \begin{align}
      A_{11}&=A_{55}=\frac{-C_{1S}(C_{24}+C_{2S})[2C_{23}^2+C_{3S}(C_{24}-C_{2S})]+C_{12}^2(C_{23}^2-C_{2S}C_{3S})}{[C_{12}^2-C_{1S}(C_{24}+C_{2S})][2C_{1S}C_{23}^2+C_{12}^2C_{3S}+C_{1S}C_{3S}(C_{24}-C_{2S})]}\\
      A_{33}&=\frac{C_{12}^2+C_{1S}(C_{24}-C_{2S})}{2C_{1S}C_{23}^2+C_{12}^2C_{3S}+C_{1S}C_{3S}(C_{24}-C_{2S})}\\    
      A_{13}&=A_{31}=A_{35}=A_{53}=-\frac{C_{12}C_{23}}{2C_{1S}C_{23}^2+C_{12}^2C_{3S}+C_{1S}C_{3S}(C_{24}-C_{2S})}\\
      A_{15}&=A_{51}=\frac{C_{12}^2(C_{23}^2+C_{24}C_{3S})}{[C_{12}^2-C_{1S}(C_{24}+C_{2S})][2C_{1S}C_{23}^2+C_{12}^2C_{3S}+C_{1S}C_{3S}(C_{24}-C_{2S})]}
  \end{align}
\end{subequations}
\end{widetext}

To represent the tunable coupler and the second qubit, we can substitute the subscript C for subscript 3 and the subscript 2 for subscript 5, respectively, in the Hamiltonian expression:
\begin{equation}
  \begin{aligned}
    H=&\sum_{j=1,\text{C},2}\left[\frac12A_{jj}Q_j^2-E_{Jj}\cos(\phi_j+\phi_{0j})\right]\\
    &+\sum_{\substack{j,k=1,\text{C},2\\k\neq j}}\frac12A_{jk}Q_jQ_k
  \end{aligned}
  \label{equ:A12}
\end{equation}
After the canonical quantization procedure, we obtain:
\begin{equation}
  \begin{aligned}
    \hat{H}=&\sum_{j=1,\text{C},2}\left[4E_{Cj}\hat{n}_j^2-E_{Jj}\cos(\hat{\phi}_j+\phi_{0j})\right]\\
    &+\sum_{\substack{j,k=1,\text{C},2\\k\neq j}}2E_{jk}\hat{n}_j\hat{n}_k
  \end{aligned}
  \label{equ:A4}
\end{equation}
where $\hat{n}_j=-i\partial/\partial\phi_j=Q_j/2e,\ j\in\{1,\text{C},2\}$ is the Cooper pair operator, $E_{Cj}=e^2A_{jj}/2$ is the charge energy, $E_{jk}=e^2A_{jk}/2,\ j,k\in\{1,\text{C},2\},\ k\neq j$, and $\hat{\phi}_j=2\pi\Phi_j/\Phi_0$ is the conjugate variable operator of $\hat{n}_j$, $\Phi_0=h/2e$. In addition, we approximate that the anharmonicity of the qubits and the tunable coupler is $\alpha_j=-E_{Cj}$.

Next, we perform a second quantization by introducing the creation operator $\hat{a}_j^\dagger$ and the annihilation operator $\hat{a}_j,\ j\in\{1,\text{C},2\}$, which satisfy the commutation relationship $[\hat{a}_j,\hat{a}_j^\dagger]=1$. This allows us to change the system to the harmonic oscillator basis \cite{Chen}:
\begin{subequations}
  \begin{align}
      \hat{n}_j&=in_j^{\text{ZPF}}(\hat{a}_j^\dagger-\hat{a}_j)\\
      \hat{\phi}_j&=\phi_j^{\text{ZPF}}(\hat{a}_j^\dagger+\hat{a}_j)
  \end{align}\label{equ:A2}
\end{subequations}
where the zero-point fluctuations are:
\begin{subequations}
  \begin{align}
      n_j^{\text{ZPF}}&=\frac{1}{\sqrt{2}}\left(\frac{E_{Jj}}{8E_{Cj}}\right)^{\frac14}\\
      \phi_j^{\text{ZPF}}&=\frac{1}{\sqrt{2}}\left(\frac{8E_{Cj}}{E_{Jj}}\right)^{\frac14}
  \end{align}\label{equ:A3}
\end{subequations}
Here we can further explain why it is reasonable to ignore the free particle variables mentioned above. For these variables, we can consider their $E_{Jj}=0$, which implies that their $n_j^{\text{ZPF}}=0$ and $\phi_j^{\text{ZPF}}\rightarrow\infty$. Substituting these values into the inverse transform of Eq.(\ref{equ:A2}) yields:
\begin{subequations}
  \begin{align}
      \hat{a}_j^\dagger&=\frac12\left(\frac{\hat{\phi}_j}{\phi_j^{\text{ZPF}}}+\frac{\hat{n}_j}{in_j^{\text{ZPF}}}\right)=+\frac12\frac{\hat{n}_j}{in_j^{\text{ZPF}}}\rightarrow+\infty\\
      \hat{a}_j&=\frac12\left(\frac{\hat{\phi}_j}{\phi_j^{\text{ZPF}}}-\frac{\hat{n}_j}{in_j^{\text{ZPF}}}\right)=-\frac12\frac{\hat{n}_j}{in_j^{\text{ZPF}}}\rightarrow-\infty
  \end{align}
\end{subequations}
The creation and annihilation operators only contain $\hat{n}_j$ and approach infinity, rendering them meaningless and preventing us from transforming them into the expected harmonic oscillator basis.

Substituting Eq.(\ref{equ:A2}-\ref{equ:A3}) into Eq.(\ref{equ:A4}) and expanding the cosine terms up to the sixth order, we can obtain:
\begin{widetext}
\begin{equation}
  \hat{H}=\sum_{j=1,\text{C},2}\left[\omega_j+\frac{E_{Cj}}{2}\left(1-\frac{5\xi_j}{18}\right)-\frac{E_{Cj}}{2}\left(1-\frac{\xi_j}{6}\right)\hat{a}_j^\dagger\hat{a}_j\right]\hat{a}_j^\dagger\hat{a}_j+\sum_{k=1,2}g_{k\text{C}}(\hat{a}_k^\dagger-\hat{a}_k)(\hat{a}_\text{C}^\dagger-\hat{a}_\text{C})+g_{12}(\hat{a}_1^\dagger-\hat{a}_1)(\hat{a}_2^\dagger-\hat{a}_2)
  \label{equ:A7}
\end{equation}
\end{widetext}
where the sixth-order corrected frequency of the qubits and the tunable coupler is:
\begin{equation}
  \begin{aligned}
    &\omega_j=\sqrt{8E_{Jj}E_{Cj}}-E_{Cj}\left(1-\frac{\xi_j}{4}\right),\\
    &\xi_j=\sqrt{\frac{2E_{Cj}}{E_{Jj}}},\ j\in\{1,\text{C},2\}
  \end{aligned}
\end{equation}
and the coupling strengths are:
\begin{subequations}
  \begin{align}
      g_{k\text{C}}&=\frac{E_{k\text{C}}}{\sqrt{2}}\left(\frac{E_{Jk}E_{J\text{C}}}{E_{Ck}E_{C\text{C}}}\right)^{\frac14}\left[1-\frac18(\xi_k+\xi_\text{C})\right],\ k\in\{1,2\}\\
      g_{12}&=\frac{E_{12}}{\sqrt{2}}\left(\frac{E_{J1}E_{J2}}{E_{C1}E_{C2}}\right)^{\frac14}\left[1-\frac18(\xi_1+\xi_2)\right]
  \end{align}\label{equ:A15}
\end{subequations}
where $g_{k\text{C}}$ is the coupling strength between the qubit and the tunable coupler and $g_{12}$ is the coupling strength between the qubits. Typically, the Josephson energies of qubits, i.e. $E_{J1}$ and $E_{J2}$, are fixed, while the Josephson energy of the tunable coupler $E_{J\text{C}}$ is tunable through an external flux $\phi_{\text{e}\text{C}}$, i.e. $E_{J\text{C}}=E_{J\text{C}}(\phi_{\text{e}\text{C}})$ (see Eq.(\ref{equ:A1})). Consequently, $g_{12}$ remains fixed, while $g_{j\text{C}}=g_{j\text{C}}(\phi_{\text{e}\text{C}})$ becomes tunable (note that $\xi_\text{C}=\xi_\text{C}(\phi_{\text{e}\text{C}})$ also depends on $\phi_{\text{e}\text{C}}$).

To obtain the effective coupling strength between the two qubits, it is necessary to decouple the tunable coupler from the system. This can be achieved using the SWT (Schrieffer-Wolff transformation) \cite{SWT}:
\begin{equation}
  \widetilde{\hat{H}}=\hat{U}^\dagger\hat{H}\hat{U}=e^{-\hat{S}}\hat{H}e^{\hat{S}}=\hat{H}-[\hat{S},\hat{H}]+\frac12[\hat{S},[\hat{S},\hat{H}]]+\ldots\label{equ:A5}
\end{equation}
The Hamiltonian can be split into two parts: $\hat{H}=\hat{H}_0+\hat{H}_c$. Here, $\hat{H}_0$ comprises the diagonal terms of the Hamiltonian along with the direct coupling term between the qubits (i.e., the term in Eq.(\ref{equ:A7}) with the coefficient $g_{12}$), while $\hat{H}_c$ represents the off-diagonal terms associated with the tunable coupler. To eliminate $\hat{H}_c$, a small $\hat{S}$ needs to be identified that satisfies $[\hat{S},\hat{H}_0]=\hat{H}_c$, which enables the approximation to the second order:
\begin{equation}
  \widetilde{\hat{H}}\approx\hat{H}_0-\frac12[\hat{S},\hat{H}_c]
\end{equation}

The unitary operator that satisfies the aforementioned condition is:
\begin{equation}
  \hat{U}=e^{\sum_{k=1,2}\left[-\frac{g_{k\text{C}}}{\Delta_k}
  (\hat{a}_k^\dagger\hat{a}_\text{C}-\hat{a}_k\hat{a}_\text{C}^\dagger)
  -\frac{g_{k\text{C}}}{\Sigma_k}(\hat{a}_k^\dagger\hat{a}_\text{C}^\dagger-\hat{a}_k\hat{a}_\text{C})\right]}\label{equ:A6}
\end{equation}
where $\Delta_k=\omega_\text{C}-\omega_k$ and $\Sigma_k=\omega_\text{C}+\omega_k,\ k\in\{1,2\}$. Typically, $g_{k\text{C}}\ll\Delta_k,\Sigma_k$, allowing us to treat $g_{k\text{C}}/\Delta_k$ and $g_{k\text{C}}/\Sigma_k$ as small quantities. With this assumption, we can substitute Eq.(\ref{equ:A6}) into Eq.(\ref{equ:A5}), approximate it to the second order, truncate it to the second energy level, apply the rotating wave approximation, and put the tunable coupler in the ground state to obtain the effective qubit-qubit coupling strength $g_{\text{eff}}$:
\begin{subequations}
  \begin{align}
      \hat{H}_\text{eff}&\approx\sum_{k=1}^2\left(-\frac12\omega_k^{\text{eff}}\hat{\sigma}_k^z\right)+g_{\text{eff}}(\hat{\sigma}_1^+\hat{\sigma}_2^-+H.C.)\\
      \omega_k^{\text{eff}}&=\omega_k-g_{k\text{C}}^2\left(\frac{1}{\Delta_k}+\frac{1}{\Sigma_k}\right),\ k\in\{1,2\}\\
      g_{\text{eff}}&=g_{12}-\frac{g_{1\text{C}}g_{2\text{C}}}{2}\sum_{k=1}^2\left(\frac{1}{\Delta_k}+\frac{1}{\Sigma_k}\right)
  \end{align}
\end{subequations}
Similarly, while $g_{12},\ \omega_1$ and $\omega_2$ are fixed, $g_{1\text{C}},\ g_{2\text{C}}$ and $\omega_\text{C}$ can be tuned using $\phi_{\text{e}\text{C}}$, resulting in a tunable effective coupling strength $g_{\text{eff}}$. Typically, the frequency of the tunable coupler is set above the frequencies of the qubits ($\omega_\text{C}>\omega_k,\ k\in\{1,2\}$), so that $\Delta_k>0$. Additionally, $g_{12},\ g_{k\text{C}}$, and $\Sigma_k$ are all positive. As a result, we can identify suitable parameters that allow the TCCP architectures to achieve the modulation of $g_{\text{eff}}$ from positive to negative within a certain range of the flux variation of the tunable coupler.

A similar analysis can be performed for the TCCP architecture with one connecting pad, as shown in FIG.\ref{fig:A1}(c-d). In this case, the kinetic and potential energies become (to distinguish it from the design with two connecting pads, we use symbols with $'$):
\begin{subequations}
  \begin{align}
      T'=&\sum_{j=1}^4\frac12C_{0j}'\dot{\Phi}_j'^2+\sum_{j,k\neq0}\frac12C_{jk}'(\dot{\Phi}_j'-\dot{\Phi}_k')^2\\
      U'=&-E_{J1}'(\phi_{\text{e}1}')\cos(\phi_1'+\phi_{01}')-E_{J\text{C}}'(\phi_{\text{eC}}')\cos(\phi_3'+\phi_{0\text{C}}')\notag\\
      &-E_{J2}'(\phi_{\text{e}2}')\cos(\phi_4'+\phi_{02}')
  \end{align}
\end{subequations}
Next, we can also get the relationship between the charge and the magnetic flux, whose final matrix form is:
\begin{equation}
  \left(\begin{matrix}
      Q_1' \\ Q_2' \\ Q_3' \\ Q_4'
  \end{matrix}\right)=\left(\begin{matrix}
      C_{1S}' & -C_{12}' & 0 & 0 \\
      -C_{12}' & C_{2S}' & -C_{23}' & -C_{24}' \\
      0 & -C_{23}' & C_{3S}' & 0 \\
      0 & -C_{24}' & 0 & C_{4S}'
  \end{matrix}\right)\left(\begin{matrix}
      \dot{\Phi}_1' \\ \dot{\Phi}_2' \\ \dot{\Phi}_3' \\ \dot{\Phi}_4'
  \end{matrix}\right)\label{equ:8}
\end{equation}
where $C_{1S}'=C_{01}'+C_{12}',\ C_{2S}'=C_{02}'+C_{12}'+C_{23}'+C_{24}',\ C_{3S}'=C_{03}'+C_{23}',\ C_{4S}'=C_{04}'+C_{24}'$. Here we can also ignore the free particle variables $\Phi_2$ and the matrix form represented by the notations becomes:
\begin{equation}
  \left(\begin{matrix}
      \dot{\Phi}_1' \\ \dot{\Phi}_3' \\ \dot{\Phi}_4'
  \end{matrix}\right)=\left(\begin{matrix}
      A_{11}' & A_{13}' & A_{14}' \\
      A_{31}' & A_{33}' & A_{34}' \\
      A_{41}' & A_{43}' & A_{44}' 
  \end{matrix}\right)\left(\begin{matrix}
      Q_1' \\ Q_3' \\ Q_4'
  \end{matrix}\right)
\end{equation}
If we assume that the design of the TCCP architecture is symmetric, i.e. $C_{12}'=C_{24}',\ C_{01}'=C_{04}'$, we can obtain:
\begin{subequations}
  \begin{align}
      A_{11}'&=A_{44}'=\frac{C_{12}'^2C_{3S}'+C_{1S}'(C_{23}'^2-C_{2S}'C_{3S}')}{C_{1S}'[2C_{12}'^2C_{3S}'+C_{1S}'(C_{23}'^2-C_{2S}'C_{3S}')]}\\
      A_{33}'&=\frac{2C_{12}'^2-C_{1S}'C_{2S}'}{C_{1S}'C_{23}'^2+2C_{12}'^2C_{3S}'-C_{1S}'C_{2S}'C_{3S}'}\\    
      A_{13}'&=A_{31}'=A_{34}'=A_{43}'\notag\\
      &=-\frac{C_{12}'C_{23}'}{2C_{12}'^2C_{3S}'+C_{1S}'(C_{23}'^2-C_{2S}'C_{3S}')}\\
      A_{14}'&=A_{41}'=\frac{C_{12}'^2C_{3S}'}{C_{1S}'(C_{1S}'C_{2S}'C_{3S}'-C_{1S}'C_{23}'^2-2C_{12}'^2C_{3S}')}
  \end{align}
\end{subequations}
For the design with one connecting pad, we can replace subscript 3 with subscript C and subscript 4 with subscript 2, and remove $'$ to obtain the same Hamiltonian as in Eq.(\ref{equ:A12}). We will then perform the same calculations as we did for the design with two connecting pads.

In addition to the above two architectures, we also consider two asymmetrical architectures shown in FIG.\ref{fig:A1}(e-f). The architecture shown in FIG.\ref{fig:A1}(e) has two connecting pads, but its tunable coupler is only capacitively coupled to one of the connecting pads. The architecture shown in FIG.\ref{fig:A1}(f) has one connecting pad, but its coupler is sandwiched between the connecting pad and one of the qubits. These architectures can be analyzed using the same method as above, so we will not repeat the details.

\section{Measurement Setups \label{sec:C}}
\setcounter{figure}{0}
\setcounter{table}{0}

\begin{figure}[htb]
  \center{\includegraphics[width=8cm]  {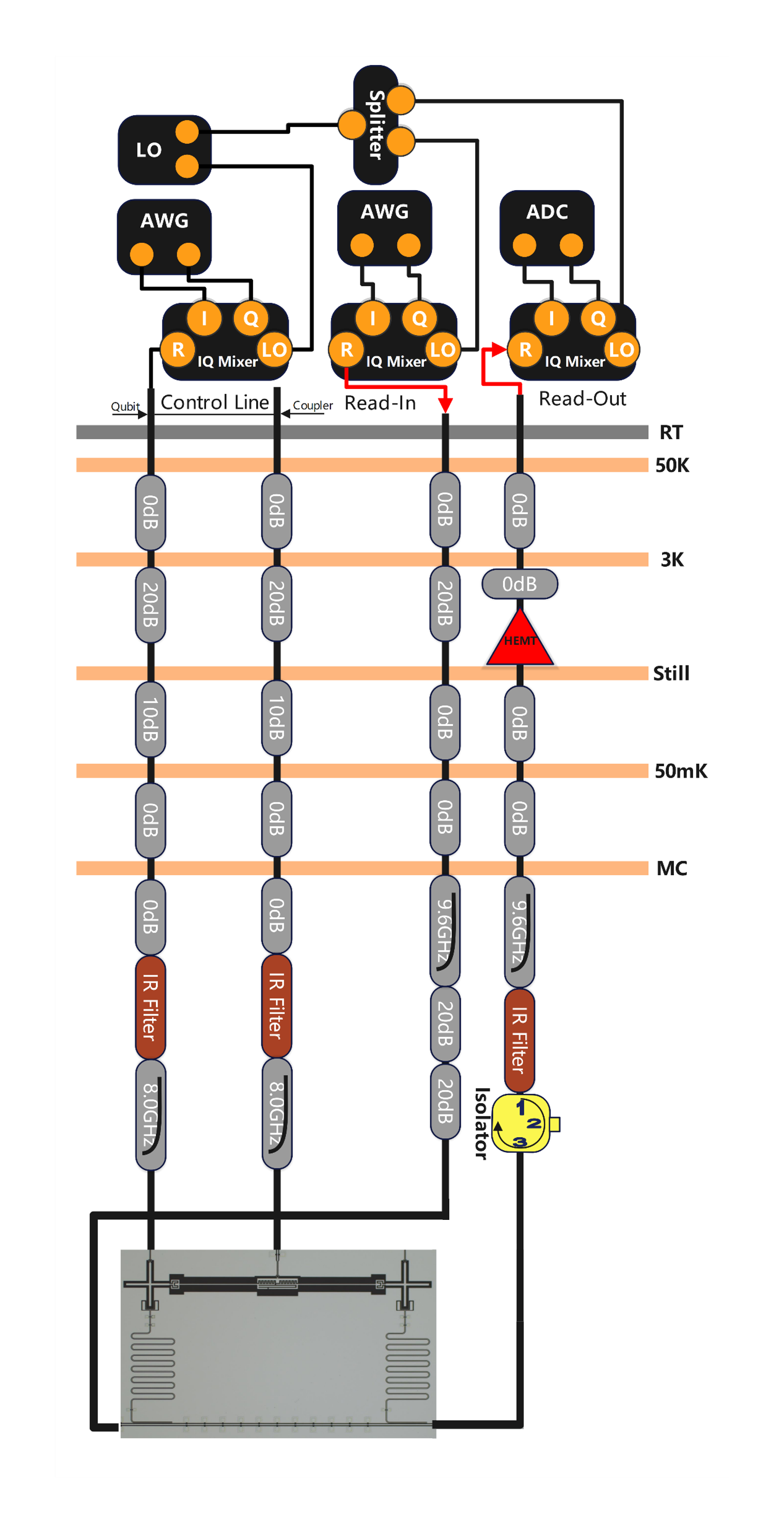}}
  \caption{A brief scheme of the experimental setups for our superconducting qubits.}
  \label{fig:C1}
\end{figure}

Our chip was mounted inside a BlueFors dilution refrigerator (DR), which had a mixing chamber plate (MC) temperature of approximately 10mK. To ensure optimal performance of all qubits and tunable couplers, we employed typical cabling plans for their driving lines. Proper attenuators and filters were set up to mitigate the Johnson-Nyquist noise and higher harmonics. The readout line, which was shared among the qubits, was equipped with an isolator and a high-electron-mobility transistor (HEMT). The pulses transmitted through all transmission lines were generated by a combination of an arbitrary waveform generator (AWG) and a local oscillator (LO). These signals were sequenced by our control system, which first sent the XY/Z control signals. The signals were then received by an analog-to-digital converter (ADC) and transferred to our control system for analysis. All the setups are shown in FIG.\ref{fig:C1}.

\section{Measurements Related to TCCP \label{sec:E}}
\setcounter{figure}{0}
\setcounter{table}{0}

The coupling strength between the qubit and coupler $g_{\text{QC}}$ is a crucial parameter that reflects the accuracy of the fabrication process and provides the fundamental information about the TCCP architectures. The method of measurement refers to some previous work of our group \cite{10QubitCoupler,BlackHole}. As we gradually move the frequency of the tunable coupler across the qubit, the qubit can no longer be excited to the $|1\rangle$ state at the previously calibrated frequency. This is due to the changes in the quantity $g_{\text{QC}}^2/\Delta_{\text{QC}}$, where $\Delta_{\text{QC}}=\omega_\text{C}-\omega_\text{Q}$, and the system is no longer in the dispersive regime as the frequency of the tunable coupler approaches the qubit. We illustrate this effect in terms of the Hamiltonian:
\begin{equation}
    H = \frac{1}{2}\omega_\text{Q} \hat{\sigma}_z \otimes\hat{\mathcal{I}}+
    \frac{1}{2}\omega_\text{C} \hat{\mathcal{I}}\otimes\hat{\sigma}_z +g_{\text{QC}}\hat{\sigma}_y\otimes\hat{\sigma}_y
\end{equation}
As a consequence of the deviation from the dispersive regime, the energy gap between the ground state and the first excited state is no longer equal to the frequency of the qubit, indicating that it is not appropriate to treat the qubit and the tunable coupler as effectively decoupled. Theoretically, since $\Delta_{\text{QC}} \gg g_{\text{QC}}$ is not applicable, the energy of the first and second excited states of the coupled system of the qubit and the tunable coupler can be approximately expressed as $(\omega_\text{Q}+\omega_\text{C}-\sqrt{\Delta_{\text{QC}}^2+4g_{\text{QC}}^2})/2$ and $(\omega_\text{Q}+\omega_\text{C}+\sqrt{\Delta_{\text{QC}}^2+4g_{\text{QC}}^2})/2$ respectively. These energies approach $\omega_\text{Q}$ and $\omega_\text{C}$ respectively, only when $\Delta_{\text{QC}}$ is sufficiently large to ignore $g_{\text{QC}}$. When we excite the system around the frequency of $\omega_\text{Q}$ by gradually reducing the frequency of the tunable coupler, an anticross between the first and second excited states occurs, giving rise to a gap of $\sqrt{\Delta_{\text{QC}}^2+4g_{\text{QC}}^2}$ between the anticross. The minimum separation between the anticross is $2g_{\text{QC}}$, which we experimentally measure as a key parameter. To determine $g_{\text{QC}}$ accurately, we sweep the variable frequency of the tunable coupler around the frequency of the qubit that we calibrated previously at the sweet point. Prior to this measurement, we perform the spectroscopy of the tunable coupler. Although there is no direct readout transmission line connected to the tunable coupler, we infer the state of the tunable coupler by measuring the $|1\rangle$ probability of one of the qubits, which is affected by $g_{\text{QC}}$. Specifically, $g_{\text{QC}}^2/\Delta_{\text{QC}}$ changes as the detuning between the qubit and the tunable coupler varies.

\begin{figure}[htb]
    \center{\includegraphics[width=8.5cm]  {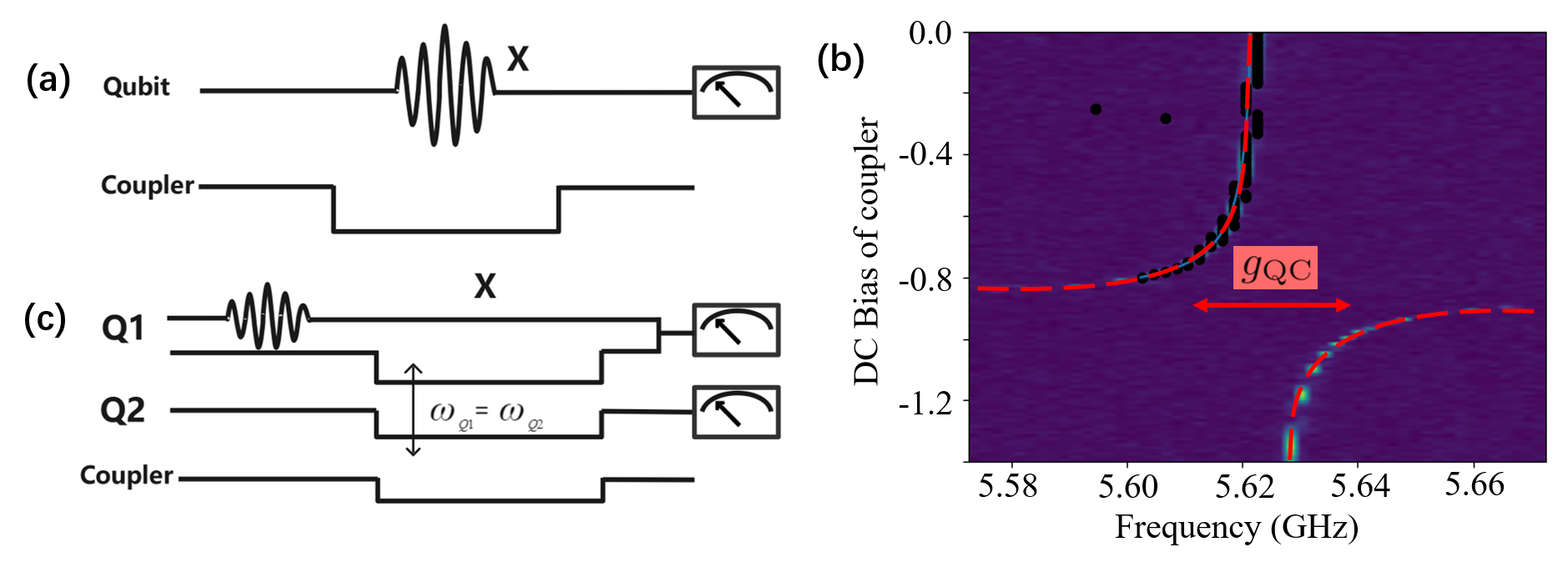}}
    \caption{(a) The pulse sequence for measuring the anticross to extract $g_{\text{QC}}$. We apply a $\pi$ pulse to a qubit and simultaneously apply a DC bias to the tunable coupler coupled to the qubit. We then measure the qubit. (b) The anticross spectrum between the qubit and the tunable coupler, with the excitation probability of the qubit to be plotted against the frequency of the $\pi$ pulse applied to the qubit. The brighter spot indicates a higher probability of the excited qubit, and the red dashed line represents the curve used to fit $g_{\text{QC}}$. (c) The pulse sequence for measuring the SWAP frequency to extract $g_{\text{eff}}$. We first excite a qubit with a $\pi$ pulse, then apply the DC bias to make the two qubits resonant and bias the tunable coupler. Finally, we measure the qubits.}
    \label{fig:E1}
\end{figure}

FIG.\ref{fig:E1}(a) shows the pulse used for measuring $g_{\text{QC}}$. When the qubit and the tunable coupler are close to resonance, we apply a $\pi$ pulse to the qubit at each DC bias of the tunable coupler, while varying the frequency of the pulse. When the frequency of the pulse match the qubit, we will observe an excitation of the qubit, indicated by a bright spot in FIG.\ref{fig:E1}(b). By fitting the resulting anticross feature, we determine $g_{\text{QC}}$. An example of the measurement result is shown in FIG.\ref{fig:E1}(b).

For measuring $g_{\text{eff}}$, we use the pulse shown in FIG.\ref{fig:E1}(c). We apply a $\pi$ pulse to one of the qubits to excite it, and then tune the both qubits to the same frequency by adjusting the DC bias. Meanwhile, we modulate the tunable coupler to different frequencies. After a delay, we measure the qubits and will observe the SWAP between them, as shown in FIG.\ref{fig:3}(a). By Fourier transforming the data in FIG.\ref{fig:3}(a), we can obtain the Fourier spectrum shown in FIG.\ref{fig:3}(b), which corresponds to $g_{\text{eff}}$. We determined $g_{\text{eff}}$ by fitting the data in FIG.\ref{fig:3}(b). In addition to the Design B mentioned in the main text, we also measured the parameters related to the tunable coupling for Design A and Design D. The maximum frequency of the tunable coupler $\omega_\text{C}$, $g_{\text{QC}}$, and the tunable range of $g_{\text{eff}}$ for each design are summarized in Table \ref{table:E1}.

\begin{table}[htb]
    \center\begin{tabular}{llll}
    \hline\hline
    Design  & $\ \ \omega_\text{C}$(GHz) & $\ \ g_{\text{QC}}$(MHz) & $\ \ \ \ \ g_{\text{eff}}$(MHz) \\\hline
    A       & $\ \ $6.852 & $\ \ $49.6 $\&$ 68.2 & $\ \ \ \ \ $-18 $\sim$ 1.5 \\
    B       & $\ \ $7.089 & $\ \ $67.6 $\&$ 84.8 & $\ \ \ \ \ $-25 $\sim$ 3 \\
    D       & $\ \ $7.131 & $\ \ $66.2 $\&$ 63.1 & $\ \ \ \ \ $-22 $\sim$ 13 \\\hline\hline
    \end{tabular}
    \center\caption{The measurement results related to TCCP for Design A, B and D.} 
    \label{table:E1}
\end{table}

\section{Residual ZZ Interaction \label{sec:F}}
\setcounter{figure}{0}
\setcounter{table}{0}

\begin{figure}[htb]
  \center{\includegraphics[width=8cm]  {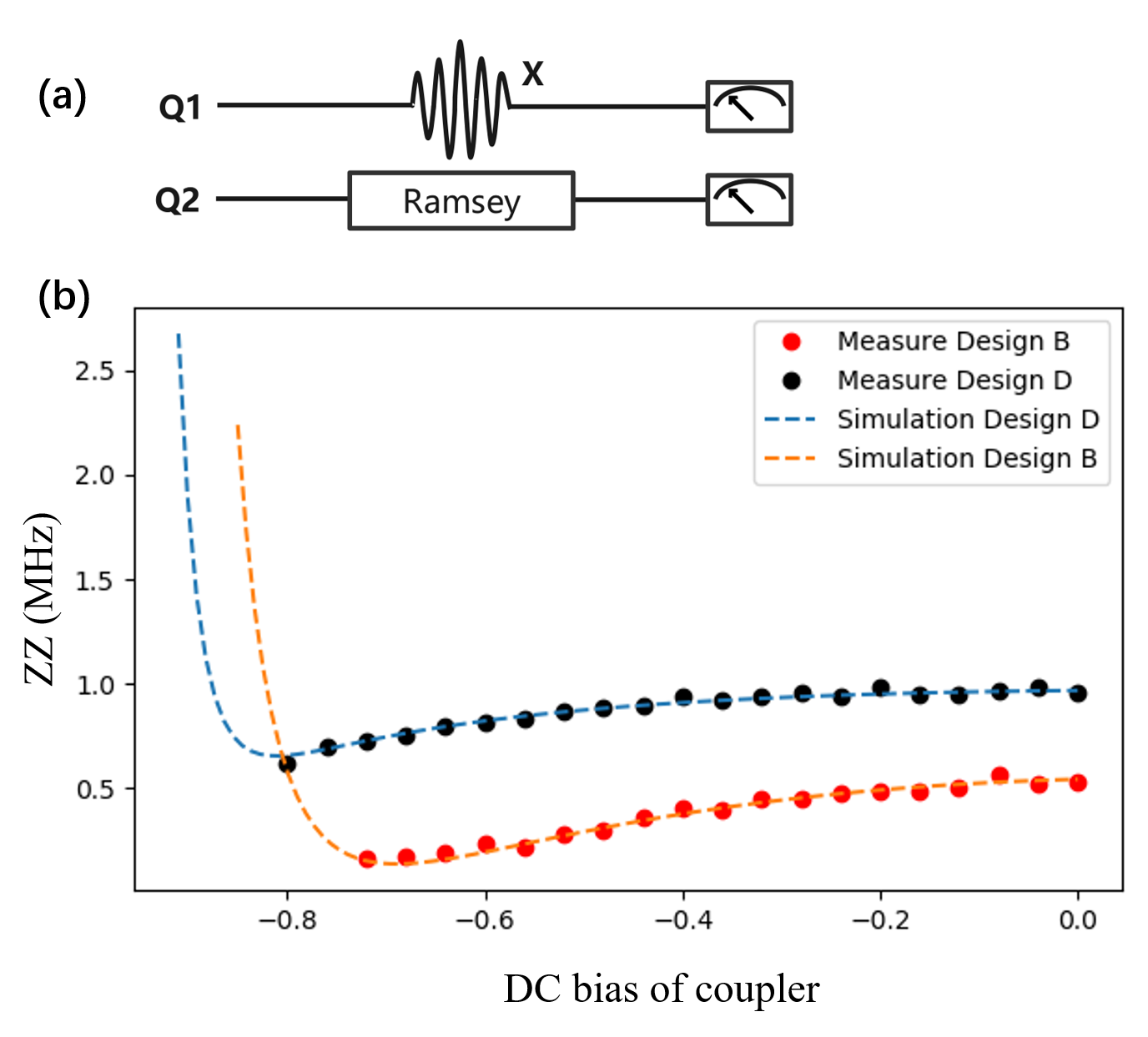}}
  \caption{(a) The pulse plot of measuring the residual ZZ interaction. Apply a $\pi$ pulse to excite one qubit and do Ramsey experiment on another qubit. (b) The simulated and Experimental image of the residual ZZ interaction of Design B and D. The experimental dates are shown as dots while the simulated results are depicted as dashed lines.}
  \label{fig:F1}
\end{figure}

In addition to the coupling strength, we also care about the residual ZZ interaction. The residual ZZ interaction can be represented simply as:
\begin{equation}
  \text{ZZ}=\omega_{|101\rangle}-\omega_{|100\rangle}-\omega_{|001\rangle}+\omega_{|000\rangle}
\end{equation}
where $\omega_{|nml\rangle}$ represents the energy of the system in the state ${|nml\rangle}$, the indexes $n,\ m$ and $l$ represent the state of the first qubit, the tunable coupler and the second qubit, respectively. We can use perturbation expansion to derive an approximate expression of the residual ZZ interaction. The second-order, third-order and fourth-order expansions are \cite{Sunluyan}:
\begin{widetext}
\begin{subequations}
  \begin{align}
      \text{ZZ}^{(2)}=&-\frac{2g_{12}^2(\alpha_1+\alpha_2)}{(\Delta_{12}-\alpha_1)(\Delta_{12}+\alpha_2)}\\
      \text{ZZ}^{(3)}=&-2g_{12}g_{1\text{C}}g_{2\text{C}}\left[\frac{1}{\Delta_2}\left(\frac{1}{\Delta_{12}}+\frac{2}{\alpha_1-\Delta_{12}}\right)+\frac{1}{\Delta_1}\left(\frac{2}{\alpha_2+\Delta_{12}}-\frac{1}{\Delta_{12}}\right)\right]\\
      \text{ZZ}^{(4)}=&\frac{g_{1\text{C}}^2g_{2\text{C}}^2}{\Delta_1^2}\left(\frac{2}{\alpha_2+\Delta_{12}}-\frac{1}{\Delta_{12}}+\frac{1}{\Delta_2}\right)+\frac{g_{1\text{C}}^2g_{2\text{C}}^2}{\Delta_2^2}\left(\frac{2}{\alpha_1-\Delta_{12}}+\frac{1}{\Delta_{12}}+\frac{1}{\Delta_1}\right)-\frac{2g_{1\text{C}}^2g_{2\text{C}}^2}{\Delta_1+\Delta_2+\alpha_\text{C}}\left(\frac{1}{\Delta_1}+\frac{1}{\Delta_2}\right)^2
  \end{align}\label{equ:A25}
\end{subequations}
\end{widetext}
where $\Delta_{12}=\omega_1-\omega_2$ and we approximate the residual ZZ interaction with $\text{ZZ}\approx\text{ZZ}^{(2)}+\text{ZZ}^{(3)}+\text{ZZ}^{(4)}$. Similarly, the residual ZZ interaction also depends on $\phi_{\text{e}\text{C}}$.

The residual ZZ interaction originates from an effective Hamiltonian in Eq.(\ref{equ:F1}), where we consider our TCCP architectures in the dispersive regime, i.e., the detuning of the qubits and the tunable coupler satisfies $\Delta_{\text{QC}} = |\omega_\text{Q}-\omega_\text{C}|\gg g_{\text{QC}}$, and approximate the higher energy level interactions to the truncation of two-level subspace, by using SWT:
\begin{equation}
  \begin{aligned}
    H=&\frac{\omega_1}{2}\hat{\sigma}_z \otimes\hat{\mathcal{I}}
      +\frac{\omega_2}{2}\hat{\mathcal{I}}\otimes\hat{\sigma}_z
      +\frac{g_{\text{eff}}}{2} \left( \hat{\sigma}_x\otimes\hat{\sigma}_x +\hat{\sigma}_y\otimes\hat{\sigma}_y\right)\\
      &+\frac{\text{ZZ}}{4}\hat{\sigma}_z \otimes \hat{\sigma}_z
  \end{aligned}\label{equ:F1}
\end{equation}
During our experimental implementation, we prepare the first qubit either on $|0\rangle$ state or $|1\rangle$ state and monitor the frequency of the second qubit by Ramsey fringe experiment as shown in FIG.\ref{fig:F1}(a). Notice that the residual ZZ interaction is the function of $\omega_\text{C}$ and $g_{12}$. We gradually tune the frequency of the tunable coupler to observe the change of the residual ZZ interaction. According to the residual ZZ interaction, we find a proper initial frequency of the tunable coupler to prepare the adiabatic CZ gate and acquire $g_{12}$ through the numerical simulation. After the experiment, we plotted the experimental data together with the curve of the numerical simulation of Design B and D as shown in FIG.\ref{fig:F1}(b-c), where the dots represent the experimental dates and the dashed lines represented simulation. The curve of the numerical simulation was obtained by substituting our measured parameters of the design into Eq.(\ref{equ:A25}) for numerical drawing. In the process of tuning the tunable coupler from its maximum frequency to near-resonance with the qubit, the residual ZZ interaction of the TCCP architectures could be lower than 1MHz. Meanwhile, $g_{12}$ of Design B and D was approximately 7.8MHz and 19MHz respectively. We found that since $g_{12}$ of Design D was larger than that of Design B, the residual ZZ interaction of Design D was larger. It is consistent with the parameters of our designs.

\nocite{*}


\begin{thebibliography}{40}
\bibitem[Pat()]{Patent}
Before writing up this manuscript, we have filed a Chinese patent for our
  design on June 23, 2022, the patent number is CN 202210717997.7.
\end{thebibliography}
\providecommand{\noopsort}[1]{}\providecommand{\singleletter}[1]{#1}%

\end{document}